\documentclass[paper=letter, fontsize=12pt]{article}
\usepackage[english]{babel} 
\usepackage{amsmath,amsfonts,amsthm} 
\usepackage[utf8]{inputenc}
\usepackage{float}
\usepackage{lipsum} 
\usepackage{blindtext}
\usepackage{graphicx} 
\usepackage{caption}
\usepackage[sc]{mathpazo} 
\usepackage[T1]{fontenc} 
\usepackage{microtype} 
\usepackage[hmarginratio=1:1,top=32mm,columnsep=20pt]{geometry} 
\usepackage{multicol} 
\usepackage{booktabs} 
\usepackage{float} 
\usepackage{hyperref} 
\usepackage{lettrine} 
\usepackage{paralist} 
\usepackage{abstract} 
\usepackage{titlesec} 
\usepackage{natbib, url}
\usepackage{subfig}
\usepackage{listings}
\usepackage{bm}
\usepackage{cancel}
\usepackage{xcolor}
\renewcommand\thesection{\Roman{section}} 
\renewcommand\thesubsection{\Roman{subsection}} 

\titleformat{\section}[block]{\large\scshape\centering}{\thesection.}{1em}{} 
\titleformat{\subsection}[block]{\large}{\thesubsection.}{1em}{} 
\usepackage{fancyhdr} 
\newcommand{\T}{\top}
\newcommand{\given}{\, | \,}
\newcommand*{\QEDB}{\hfill\ensuremath{\square}}%
\newtheorem{example}{Example}

\title{\vspace{-15mm}\fontsize{24pt}{10pt}\selectfont\textbf{Bayesian Inference for Finite Populations Under Spatial Process Settings}} 

\author{
\large
{\textsc{Alec M. Chan-Golston}}\\[2mm]
{\textsc{UCLA Department of Biostatistics }}\\[2mm]
\normalsize \href{mailto:aleccg@ucla.edu}{aleccg@ucla.edu}\\[2mm] 
\large
{\textsc{Sudipto Banerjee}}\\[2mm]
{\textsc{UCLA Department of Biostatistics }}\\[2mm]
\normalsize \href{mailto:sudipto@ucla.edu}{sudipto@ucla.edu}\\[2mm] 
\large
{\textsc{Mark S. Handcock}}\\[2mm]
{\textsc{UCLA Department of Statistics }}\\[2mm]
\normalsize \href{mailto:handcock@stat.ucla.edu}{handcock@stat.ucla.edu}\\[2mm] 
}
\date{November 20, 2019}
\providecommand{\keywords}[1]{\textbf{\textit{Key words:}} #1}
\begin{document}
\maketitle 
\thispagestyle{fancy} 

\label{firstpage}

\begin{abstract}
We develop a Bayesian model-based approach to finite population estimation accounting for spatial dependence. Our innovation here is a framework that achieves inference for finite population quantities in spatial process settings. A key distinction from the small area estimation setting is that we analyze finite populations referenced by their geographic coordinates. Specifically, we consider a two-stage sampling design in which the primary units are geographic regions, the secondary units are point-referenced locations, and the measured values are assumed to be a partial realization of a spatial process. Estimation of finite population quantities from geostatistical models do not account for sampling designs, which can impair inferential performance, while design-based estimates ignore the spatial dependence in the finite population. We demonstrate using simulation experiments that process-based finite population sampling models improve model fit and inference over models that fail to account for spatial correlation. Furthermore, the process based models offer richer inference with spatially interpolated maps over the entire region. We reinforce these improvements and demonstrate scaleable inference for ground-water nitrate levels in the population of California Central Valley wells by offering estimates of mean nitrate levels and their spatially interpolated maps.
\end{abstract}
\newpage

\section{Introduction}
Finite population survey sampling concerns statistical modeling and inference on finite populations from sampling designs; see, for example, \cite{Cochran:1977un}, \cite{Hartley:1975uw}, \cite{Royall:1970tf}, and \cite{Horvitz:tp}. In this article, we will concern ourselves with Bayesian inference for finite populations when the sampling units are spatially oriented. For instance, one may consider estimating the total biomass in a forest given a sample of trees, the average income of a city given a sample of individuals and their addresses, or the total amount of air pollution attributable to cars on a freeway given a sample of pollution measurements. Additionally, these techniques could be used to conduct Bayesian inference for finite population survey sampling is discussed in great detail in \cite{Gelman:2007bh}, \cite{Little:2004jz}, \cite{Ghosh:1997tp}, and \cite{Ericson:1969wi}. In this domain, there is a substantial literature on small area estimation for regionally aggregated data \citep[see, e.g.,][]{Rao:vz, Ghosh1998:kn, Ghosh1994:sa, Clayton1987:eb}, where interest lies in modeling dependencies across regions. 

Unlike the aforementioned literature on small area estimation, where the sampling units are regions such as counties, states or census-tracts, spatial process models consider quantities that, at least conceptually, exist in continuum over the entire domain. The process assigns a probability law to an uncountable subset within a $d$-dimensional Euclidean domain. In general, spatial process modeling (\citealt{Banerjee:2014wm}; \citealt{Cressie:2011uu}; and \citealt{Ripley:2004ut}) follows the generic paradigm
\begin{equation}\label{eq: generic_paradigm}
[\mbox{data}\given \mbox{process}] \times [\mbox{process}\given \mbox{parameters}] \times [\mbox{parameters}]\;,
\end{equation}
which accommodates complex dependencies and multiple sources of variation.

With regard to finite population sampling in spatial process settings, the literature appears to be considerably more scant than small area estimation. Here, \cite{VerHoef:2002xy} discuss connections between geostatistical models and classical design-based sampling and develop methods for executing model-based block kriging. \cite{Cicchitelli:2012me} present a spline regression model-assisted, design-based estimator of the mean for use on a random sample from both finite and infinite spatial populations. A linear spatial interpolator is used by \cite{Bruno2013:fb} to create a design-based predictor of values at unobserved locations which outperforms non-spatial predictors. While related to these developments, none of these techniques have presented a Bayesian approach. In this manuscript, we pursue a fully Bayesian, model-based approach as in (\ref{eq: generic_paradigm}) and carry out inference on the finite population quantities and the spatial process. 

Bayesian finite population survey sampling is essentially model-based (see, e.g., \citealt{Little:2004jz}). The population units are themselves assumed to be endowed with a probability distribution. In a Gaussian setting, \cite{Scott:1969fp} devised Bayesian hierarchical models for inferring with two-stage designs, while \cite{Malec:1985tr} extended this framework to general multi-stage (more than two-stages) models and also discussed handling unknown variances. Our current contribution focuses on incorporating survey sampling designs within (\ref{eq: generic_paradigm}).  We extend this framework to spatial process settings under the context of ignorable sampling designs (\citealt{Rubin:1976we}; \citealt{Sugden:1984tw}), where the probability of element selection is assumed independent of the measured outcome given the design variables. We specifically develop the distribution theory and algorithms for implementing (\ref{eq: generic_paradigm}) in the context of two-stage designs that encompass simple random, cluster and stratified sampling \citep[as defined in][]{Cochran:1977un} as special cases. Extension of this work to multi-stage present no new methodological difficulties, building upon \cite{Malec:1985tr}.  

The remainder of this paper evolves as follows. In Section~\ref{sec: bayesian_multi_stage_sampling}, we review a general framework for Bayesian modeling for multi-stage sampling and how simple, two-stage, and stratified random sampling designs arise as special cases. Section~\ref{sec: bayesian_spatial_process_modeling} presents modeling strategies for spatially correlated data sampled using a two-stage design, the implementation of which, along with the model proposed by \cite{Scott:1969fp}, is discussed in Section~\ref{sec:models} using Bayesian exact and Markov chain Monte Carlo (MCMC) sampling. These models are then applied to simulated data in Section~\ref{sec:sim} and then used in an analysis of nitrate levels in California groundwater in Section~\ref{sec:data_analysis}. The paper concludes with a brief discussion of the results in Section~\ref{sec:Discussion}.

\section{Bayesian modeling of multi-stage sampling}\label{sec: bayesian_multi_stage_sampling}

Suppose that $n$ samples are randomly drawn from a finite population of size $N$, $n \le N$ and for the $i$-th sampled unit, the outcome $y_i$ is measured. Without loss of generality, let the set of outcomes from the finite population be stacked into a vector $y = [y_s^\T\colon y_{ns}^\T]^\T$, where $y_s = [y_1, \dots, y_n]^\T$ and $y_{ns} = [y_{n+1}, \dots, y_N]^\T$ are vectors of outcome values from the sampled and nonsampled units, respectively. This vector has corresponding design matrix $X = [X_s^\T \colon X_{ns}^\T]^\T$, which observed for the entire finite population and denotes group membership. 

From a superpopulation perspective, we consider the finite population to be a random sample of size $N$ from an infinitely large population. This superpopulation is assumed to follow a Gaussian distribution with mean $\nu$ and a covariance function defined by parameters $\theta$. In general, we can construct the following linear regression model:
\begin{equation}\label{eq:bayes_regression}
\begin{bmatrix} y_s \\ y_{ns} \end{bmatrix} = \begin{bmatrix} X_s \\ X_{ns} \end{bmatrix} \beta+ \begin{bmatrix} \epsilon_s \\ \epsilon_{ns} \end{bmatrix} \; ;\\
\begin{bmatrix} \epsilon_s \\ \epsilon_{ns} \end{bmatrix} \sim N \Bigg( \begin{bmatrix} 0 \\ 0 \end{bmatrix} , \begin{bmatrix} V_s(\theta) & V_{s,ns}(\theta)\\ V_{ns,s}(\theta) & V_{ns}(\theta) \end{bmatrix} \Bigg) \;.
\end{equation}
Bayesian specifications further model $\beta \sim N(A\nu,V_\beta)$, where $A$ and $\nu$ are a vector with length equal to the number of groups and a scalar, respectively, and $V_\beta$ is the variance of $\beta$. The hierarchy continues with probabilistic specifications on $\nu$ and $\theta$.

Our goal is to estimate linear finite population quantities of the form $\alpha^\top y$, where $\alpha$ is a given, fixed vector of weights defined for the entire population. Suppose $\nu \sim \text{N}(0,\gamma^2)$. Define $V_{\beta\given y_s} = [(\gamma^2AA^\T + V_\beta)^{-1} + X_s^\T V_s^{-1} X_s]^{-1}$ and $Q = X_{ns} - V_{ns,s}V_s^{-1}X_s$. Fixing the variance parameters, the posterior expectation of the finite population quantity is:
\[
\text{E}[\alpha^\T y \given y_s] = \{\alpha_s^\T  +  \alpha_{ns}^\T [V_{ns,s} + \alpha_{ns}^\T QV_{\beta \given y_s}X_s^\T] V_s^{-1} \}y_s \;.
\]
Defining $B_V = V_{ns,s} + QV_{\beta\given y_s}X_s^\T$, the variance of this expectation is:
\[
\text{Var}[\text{E}[\alpha^\T y \given y_s]] = \alpha_s^\T V_s \alpha_s + 2 \alpha_s^\T B_V^\T\alpha_{ns} + 
\alpha_{ns}^\T B_V V_s^{-1}B_V^\T \alpha_{ns}\;.
\]
Additionally, the posterior variance of $\alpha^\T y$ is:
\[
\text{Var}[\alpha^\T y \given y_s] = \alpha_{ns}^\T (QV_{\beta\given y_s}Q^\T + V_{ns} -V_{ns,s}V_s^{-1}V_{s,ns})\alpha_{ns} \;.
\]

For the special case of a census, in which all members of the population are sampled, e.g. $y_s = y$, the conditional expectation of the finite population quantity is finite population consistent, in the sense that $\text{E}[\alpha^\top y \given y_s] = \text{E}[\alpha^\top y \given y] = \alpha^\top y $. Different sampling designs can incorporated by appropriately structuring the sampled and nonsampled elements. We provide a few examples below. All derivations can be located in the supplementary materials.

\begin{example}\label{ex.srs} 
Simple Random Sampling

\normalfont In simple random sampling, $n$ units are randomly drawn from a population of size $N$, where each unit in the population is independent and identically distributed with mean $\mu$ and variance $\sigma^2$. To express this as in (\ref{eq:bayes_regression}), define $y_s = [y_1, \dots, y_n]^\T$ and $y_{ns} = [y_{n+1}, \dots, y_N]^\T$, with corresponding design matrices $X_s = 1_n$ and $X_{ns} = 1_{N-n}$, respectively, where $1_n$ represents the $n \times 1$ vector of ones. Let $A=1$ and take $\beta$ to be a scalar $\mu$ with mean $\nu=0$ and variance $V_\beta = \xi^2$. Additionally, let $V_s = \sigma^2 I_n$, $V_{ns} = \sigma^2 I_{N-n}$, and $V_{s,ns} = V_{ns,s}^\T = O$, where $O$ is a matrix of zeroes of appropriate order. Define finite population weights $\alpha = [\alpha_1, \dots, \alpha_N]^\T$. Fixing the variance parameters, the posterior expectation of $\alpha^\T y$ is:
\begin{equation}
\text{E}[\alpha^\T y \given y_s] = \sum_{i=1}^n\Bigg(\alpha_i + \frac{\sum_{i=n+1}^N \alpha_i} {\frac{\sigma^2}{\xi^2} + n} \Bigg) y_i
\end{equation}
\QEDB
\end{example}

\begin{example}\label{ex.2st.rs} 
Two-Stage Sampling

\normalfont
In a more complex case, suppose that the population is divided into $N$ distinct groups defined by geography or other characteristics, with the $i$-th group of size $M_i$. Assume that within the $i$-th group, each unit is independent and identically distributed with mean $\mu_i$ and variance $\sigma_i^2$. The group means $\mu_1, \dots, \mu_N$ are independent and follow a normal distribution centered at $\nu$ with a variance of $\delta^2$, hence $\beta = \mu = [\mu_1, \dots, \mu_N]^\T$, $A = 1_N$ and $V_\beta = \delta^2I_N$. Suppose that only $n$ of the $N$ groups are randomly sampled, where $n \le N$. Without loss of generality, take the first $n$ groups to be sampled, and then within the chosen $i$-th group, $m_i$ units are randomly selected; $m_i \le M_i$, $i=1, \dots, n$. As $N-n$ groups are not sampled, the number of observed units in these groups are zero, e.g. $m_i = 0$, $i = n+1, \dots, N$. Hence, we can define the number of sampled units as $k = \sum_{i=1}^N m_i = \sum_{i=1}^n m_i$, the number of unsampled units as $K = \sum_{i = 1}^N (M_i - m_i)$, and the population total to be $T = K + k = \sum_{i=1}^N M_i$.

To examine (\ref{eq:bayes_regression}) in the context of a two-stage design, define outcome vectors $y_s= [y_1^\T , \dots, y_n^\T]^\T$ and $y_{ns}= [y_1^{*\T}, \dots, y_N^{*\T} ]^\T$, with $i^{\text{th}}$ components $y_i = [y_{i1}, \dots, y_{im_i}]^\T$ and $y_i ^*= [y_{im_i+1}, \dots, y_{iM_i}]^\T$, respectively.
The design matrix for the sampled units can be modified by fixing the $k \times N$ matrix $X_s =[ \oplus_{i = 1}^n 1_{m_i} \colon O]$, reflecting that $N-n$ of the $N$ sites are unobserved. Similarly, for the unobserved units, define $X_{ns}$ as a block diagonal, $K\times N$ matrix with upper block $[\oplus_{i = 1}^n 1_{M_i-m_i}]$ and lower block $[\oplus_{i = n+1}^{N} 1_{M_i-m_i}]$. 
For notational convenience, we also define $X_{s1} =  [\oplus_{i = 1}^n 1_{m_i}]$ and divide the group mean vector $\mu$ into sampled, $\mu_s = [\mu_1, \dots, \mu_n]^\T$, and nonsampled, $\mu_{ns} = [\mu_{n+1}, \dots, \mu_N]^\T$, components such that $\mu = [\mu_s^\T, \mu_{ns}^\T]^\T$. Note that distributional mean of $y_s$, $X_s1_N \mu$, can be simplified to $X_{s1} 1_n \mu_s$, as the mean of the sampled units does not depend on $\mu_{ns}$. Define the sampled and nonsampled covariance matrices to be $V_s = V_s^{(\sigma)} = [\oplus_{i = 1}^n \sigma_i^2 I_{m_i}]$ and $V_{ns} =V_{ns}^{(\sigma)} = [\oplus_{i = 1}^N \sigma_i^2 I_{M_i-m_i}]$, respectively, and set $V_{s,ns} = V_{ns,s}^\T = O$. Additionally, define $V^{(\sigma)} = \begin{bmatrix} V_s^{(\sigma)} & O \\ O & V_{ns}^{(\sigma)} \end{bmatrix}$.

 To make this model fully Bayesian, let  $\nu \sim N(0,\gamma^2)$ and $\delta^2 \sim IG(a,b)$. As our interest lies in estimating $\alpha^Ty$, we can derive the posterior distributions of $p(\delta^2\given y_s)$ and $p(\nu\given y_s)$ for exact sampling of the superpopulation parameters, the details of which are provided in Section~\ref{sec:exact_models}. This approach yields results similar to those derived by \cite{Scott:1969fp}, but has the added strength of including a prior distribution on $\nu$, and \cite{Ghosh:1997tp}, who replaced distributional assumptions with the assumption of posterior linearity and fixed the variance parameters.

In the two-stage case, for a set of weights $\alpha = [\alpha_{11}, \dots, \alpha_{NM_N}]^\T$, define the group mean of sampled units as $\bar{y}_i = \frac{1}{m_i}\sum_{j=1}^{m_i} y_{ij}$, $i = 1, \dots, n$, and the group weight of nonsampled units as $\alpha_i = \sum_{j=m_i+1}^{M_i} \alpha_{ij}$, $i = 1, \dots, N$. Also, let $\tilde{\gamma}^2 = \gamma^2 / \delta^2$ and define $\lambda_i = \delta^2/(\delta^2 + \sigma_i^2/m_i)$ if $i \in \{1, \dots, n\}$ and $\lambda_i = 0$ if $i \in \{n+1, \dots, N\}$. Fixing all variance parameters, the expected value of the finite population estimate is
\begin{equation}\label{eq:ss_fp}
\text{E}[\alpha^\T y \given y_s] = \sum_{i=1}^n \sum_{j=1}^{m_i} \Bigg(\alpha_{ij} + \Bigg[\alpha_i + \frac{\sum_{i=1}^N \alpha_i (1 - \lambda_i) } {1/\tilde{\gamma}^2 + \sum_{i=1}^n \lambda_i} \Bigg]\frac{\lambda_i}{m_i} \Bigg) y_{ij} \\
\end{equation}
Additionally, the two-stage case can be extended to a three-stage case by assuming that the $j^{\text{th}}$ element of the $i^{\text{th}}$ group has $m_{ij}^*$ subelements. \cite{Malec:1985tr} derive posterior distributions for the means for a three-stage sampling scheme and provide a framework to extend this to data with $t$ stages of sampling.
\QEDB
\end{example}

\begin{example}\label{ex.strt.rs} 
Stratified Random Sampling

\normalfont Stratified sampling is a special case of two-stage sampling where all groups are sampled (e.g. $n=N$ and $m_i > 0$, $i = 1, \dots, N$), and, therefore, considering the same population described in Example~\ref{ex.2st.rs}, the number of sampled units is $k = \sum_{i=1}^N m_i$, the number of nonsampled units is $K = \sum_{i=1}^N (M_i - m_i)$, and the population total is again $T = K + k = \sum_{i=1}^N M_i$. Thus, to express this design as (\ref{eq:bayes_regression}), let $y_s= [y_1^\T, \ldots, y_N^\T]^\T$ and $y_{ns}= [y_1^{*\T}, \ldots, y_N^{*\T} ]^\T$, with $i^{\text{th}}$ components, $y_i = [y_{i1}, \dots, y_{im_i}]^\T$ and $y_i ^*= [y_{im_i+1}, \dots, y_{iM_i}]^\T$, respectively. To reflect a membership to one of $N$ groups, we take $X_s =  [\oplus_{i = 1}^N 1_{m_i}]$, $X_{ns} = [\oplus_{i = 1}^N 1_{M_i - m_i}]$, $\beta = \mu = [\mu_1, \dots, \mu_N]^\T$, and $A = 1_N$. The variance components also reflect this and are defined as $V_\beta = \delta^2I_N$, $V_s = [\oplus_{i = 1}^N \sigma_i^2 I_{m_i}]$, $V_{ns} = [\oplus_{i = 1}^N \sigma_i^2 I_{M_i - m_i}]$, and $V_{s,ns} = V_{ns,s}^\T = O$. 

The posterior expectation of $\alpha^\T y$ is given by (\ref{eq:ss_fp}), noting that $n = N$ and $\lambda_i = \frac{\delta^2}{\delta^2 + \sigma_i^2 / m_i}$, $i = 1, ..., N$, is well-defined as $m_i > 0$ for all $i$. In fact, if non-informative priors are taken for the  means, e.g. $\gamma^2 \to \infty$ and $\delta^2 \to \infty$, then $\lambda_i \to 1$, $i = 1, \dots, n$ and the stratified finite population mean is $\displaystyle \text{E}\left[\frac{1}{T}1_T^\T y \given y_s\right] =\sum_{i=1}^N \frac{M_i}{T} \bar{y}_i$, \citep[see, e.g.][]{Little:2004jz}. 
\QEDB
\end{example}

\section{Bayesian spatial process modeling for multi-stage sampling}\label{sec: bayesian_spatial_process_modeling}

Data believed to be correlated as a function of geographic distance is typically described using a spatial process model. The data is assumed to be a partial realization of a Gaussian process with dependencies between elements defined by an isotropic covariance function, $C(d)$, where $d$ is the distance between any two points. Several choices for $C(d)$ are available \citep[see, e.g.][]{Banerjee:2014wm}, but a versatile family is the Mat\'ern, defined as $C(d_{ab}) = \sigma^2 + \tau^2$ if $d_{ab}= 0$ and $C(d_{ab}) =  \tau^2 \frac{2^{1-\eta}}{\Gamma(\eta)}\big(\sqrt{2\eta}d_{ab}\phi\big)^\eta K_\eta \big(\sqrt{2\eta}d_{ab}\phi\big)$ if $d_{ab} > 0$, where $K_{\eta}(\cdot)$ is the modified Bessel function, $d_{ab}$ is the distance between two locations $\ell_a$ and $\ell_b$. Here $\sigma^2$ captures variation due to measurement error or micro-resolution variation, $\tau^2$ is the spatial variance, $\phi$ is a decay parameter which determines the rate of decline in spatial association, and $\eta$ is a smoothness parameter. The exponential covariance function is a special case of Mat\'ern when $\eta = 1/2$. In this specific instance, the decay parameter is used to calculate the effective spatial range, which is the distance where spatial correlation between two points drops below 0.05. 

Extending Example~\ref{ex.2st.rs} to a geographic context, our spatial domain comprises $N$ regions. Let $\ell_{ij}$ denote the $j$-th location in region $i$. The finite population is described by values $y(\ell_{ij})$, $i=1,\ldots,N$ and $j=1,\ldots,M_i$. Let $y_s$ be the $k\times 1$ vector corresponding to measurements from the sampled locations and $y_{ns}$ be the $K\times 1$ vector of unsampled measurements. Consider the following spatial regression model for the two-stage finite population, 
\begin{equation}\label{eq:sp}
y(\ell_{ij}) = \mu(\ell_{ij}) + \omega(\ell_{ij}) + \epsilon(\ell_{ij}) \;; 
\omega \sim N(0, \Omega) \;; \;
\epsilon \sim N(0, V^{(\sigma)}) \;,
\end{equation}
where $\mu(\ell_{ij})$ is the mean of the outcome at $\ell_{ij}$, $\omega = [\omega_s^\T : \omega_{ns}^\T]^\T$ and $\epsilon= [\epsilon_s^\T: \epsilon_{ns}^\T]^\T$ are $T\times 1$ vectors formed by stacking up $\omega(\ell_{ij})$'s and $\epsilon(\ell_{ij})$'s, respectively (analogous to $y$ in Example~2). Here, $\omega$ accounts for spatial effects and $\Omega$ is the $T\times T$ spatial covariance matrix constructed with $C(d_{ab})$ and is partitioned as $\Omega = \begin{bmatrix} \Omega_s & \Omega_{s,ns} \\ \Omega_{ns,s} & \Omega_{ns} \end{bmatrix}$. Introducing spatial effects in Example~2 yields $\mu(\ell_{ij})=\mu_i$, $V_s = \Omega_s + V_s^{(\sigma)}$, $V_{ns} = \Omega_{ns} + V_{ns}^{(\sigma)}$, and $V_{s,ns} = V_{ns,s}^\T = \Omega_{s,ns}$ in (\ref{eq:bayes_regression}). This also accommodates spatial versions of Examples\ref{ex.srs}~and~\ref{ex.strt.rs} by setting $N=1$ and $N=n$, respectively.

Analogous to (\ref{eq:ss_fp}), the posterior estimate of a linear function of the population values is
\begin{equation} \label{eq:sp2_fp}
\begin{split}
\text{E}[\alpha^\T y \given y_s] 
&= \sum_{i=1}^n \sum_{j=1}^{m_i} \Bigg(\alpha_{ij}+  \alpha_{ns}^\T\Bigg\{\Omega_{ns,s}Q_s+ 
 \; [X_{ns} - \Omega_{ns,s}Q_s^{-1}X_s] \times \begin{pmatrix}\frac{1}{\delta^2} I_N + X_s^\T Q_s^{-1}X_s\end{pmatrix}^{-1} \\
 & \quad \; \times\Bigg[X_s^\T Q_s^{-1} +  \frac {\frac{1}{\delta^2} 1_N1_N^\T X_s^\T(\delta^2 X_s X_s^\T +Q_s)^{-1}}{\frac{1}{\gamma^2} + \sum_{i=1}^n \lambda_i^*}\Bigg]\Bigg\}q_{ij}\Bigg)y(\ell_{ij}) \;,
\end{split}
\end{equation}
where $\lambda^{*\T} = [\lambda_1^* \dots, \lambda_N^*] = 1_N^\T X_s^\T(\delta^2 X_s X_s^\T +\Omega_s + V_s^{(\sigma)})^{-1}X_s$, $Q_s = \Omega_s + V_s^{(\sigma)}$, and $q_{ij}$ is a set of $k$ indicator vectors of length $k$, $i = 1, \dots, n$, $j = 1, \dots, m_i$. For $i = 1$, $q_{1j}= j$ and 0 elsewhere, and if $i > 1$, $q_{ij}$ is 1 at element $\sum_{i=1}^{i-1} m_i + j$ and 0 elsewhere. This two-stage spatial model, (\ref{eq:sp}), can be written as an intercept-only spatial model by setting $\mu(\ell_{ij}) = \mu$ and $\sigma_i^2 = \sigma^2$, $i=1, \dots, N$, i.e., simplifying $V^{(\sigma)}$ to $\sigma^2 I_T$. As region is not accounted for, the design matrices $X_s$ and $X_{ns}$ are replaced with $1_k$ and $1_K$, respectively.

However, as the size of the finite population, $T$,  grows, the scaleability of (\ref{eq:sp}) diminishes due to an increased computational burden stemming from the inversion of the $T \times T$ matrix $\Omega$. To address this, we also consider a more computationally efficient model which also allows for region specific means, but specifies that each region is defined by its own process parameters and is independent from all other regions. To reflect this regional independence, we specify the covariance function specifying the spatial process $\omega(\ell)$ in (\ref{eq:sp}) to be $0$ for any two points in different regions, and equal to the value of the Mat\'ern covariance function for any two points within the same region.

Comparing the finite populations estimates given in (\ref{eq:ss_fp}) and (\ref{eq:sp2_fp}), it is evident that accounting for spatial variation results in a more complex equation, as all observed and unobserved outcome values in a population can no longer be assumed to be independent conditional on the group means. This can also be seen in the calculation of the $\lambda$ parameters, which in the two-stage model, are a simple ratio of variances. In the spatial case, however, the complexity of the parameters is increased by the addition of the spatial covariance matrix.

\section{Model Implementation and Assessment}\label{sec:models}
\subsection{General framework} \label{sec:mcmc}

A Bayesian linear model corresponding to the likelihood of the sampled data in ({\ref{eq:bayes_regression}) is
\begin{equation}\label{eq:general_regression}
p(\theta, \nu, \beta\given y_s)  \propto p(\theta)\times N(\nu\given 0, V_{\nu})\times N(\beta \given A\nu, V_\beta) 
  \times N(y_s \given X_s\beta, V_s(\theta)) \;.
\end{equation}
We use Markov chain Monte Carlo algorithms \citep[see, e.g.][]{Robert:2004fj} for sampling from (\ref{eq:general_regression}). Subsequeny Bayesian inference for $y_{ns}$ is available in posterior predictive fashion by drawing samples from 
\begin{equation}\label{eq:bayes_predict}
p(y_{ns} \given y_s) = \int p(y_{ns}\given y_s, \theta, \nu, \beta) 
 \times p(\theta, \nu, \beta \given y_s)\;  d\theta \;d\nu \;d\beta\;.
\end{equation}
Using the conditional independence of parameters in (\ref{eq:bayes_regression}), we obtain $p(y_{ns}\given y_s, \theta, \nu, \beta) = N(y_{ns}\given \mu_{ns\given s}, V_{ns\given s})$, where $\mu_{ns\given s} = X_{ns}\beta + V_{ns,s}(\theta)V_{s}(\theta)^{-1}(y_s - X_s\beta)$ and $V_{ns\given s} = V_{ns}(\theta) - V_{ns,s}(\theta)V_{s}(\theta)^{-1}V_{s,ns}(\theta)$. 
Therefore, sampling from (\ref{eq:bayes_predict}) is achieved by drawing one $\{\beta,\theta\}$ from (\ref{eq:general_regression}) followed by one $y_{ns} \sim N(X_{ns}\beta, V_{ns}(\theta))$, for each posterior sample of $\{\beta,\theta\}$. The resulting samples provide inference on the nonsampled group means $\mu_{ns}$ and Bayesian imputation for the nonsampled population units, $y_{ns}$.

These samples from the posterior predictive distribution can be used to obtain posterior finite population estimates of the form $\alpha^\T y$. We consider four models using (\ref{eq:general_regression}).

\textbf{Model 1. Two-Stage} \\
For the model provided in Example \ref{ex.2st.rs}, we take $\theta = [\gamma^2, \delta^2, \sigma_1^2, \dots, \sigma_N^2]^\T$, $p(\theta) = IG(\gamma^2\given a_\gamma, b_\gamma) \times IG(\delta^2 \given a_\delta, b_\delta) \times \prod_{i=1}^NIG(\sigma_i^2 \given a_{\sigma_i}, b_{\sigma_i})$ in (\ref{eq:general_regression}), and follow the other specifications as in the two-stage setting in Example \ref{ex.2st.rs}.

As the priors have been chosen to be fully conjugate, one can derive the full posterior conditional distributions for each of the parameters. Specifically, the variance parameters will have posterior distributions of the form $IG(a^*, b^*)$, while the rest of the parameters will have posterior distributions of the form $N(Mm, M)$. However, as only $n$ of the $N$ groups are observed, the variance terms of the unsampled groups, $\sigma_{n+1}^2, \dots, \sigma_N^2$, must either be fixed or given informative priors. If not, draws from the posterior predictive distribution corresponding to units in the nonsampled groups will have arbitrary variability and could spuriously dominate the finite population estimates. 

\textbf{Model 2. Spatial} \\
Under (\ref{eq:general_regression}), the intercept-only spatial model defines $\theta = [\phi, \delta^2, \sigma^2, \tau^2]^\T$ with corresponding prior distribution $p(\theta) = p(\phi) \times IG(\delta^2\given a_\delta, b_\delta) \times IG(\sigma^2\given a_\sigma, b_\sigma) \times IG(\tau^2\given a_\tau, b_\tau)$ and $V_\beta = \delta^2$, $V_s = \Omega_s + \sigma^2 I_k$. As there are no group terms, replace $X_s$ with $1_k$ and take $\nu = 0$ with probability 1, e.g. $V_\nu^{-1} = 0$.

Unlike model 1, regardless of the prior distribution placed on $p(\phi)$, a closed-form posterior distribution cannot be found for $\phi$. In practice, $\phi$ is often fixed using an estimate found from a variogram and then full posterior conditional distributions can be found using the same techniques described for the non-spatial case. However, MCMC can still be implemented by specifying a prior distribution for $\phi$ (\citealt{Banerjee:2014wm}), which is often taken to be a uniform distribution. 

To recover the spatial effects $\omega$ absorbed into the variance parameter of $y$, note that $\beta = \mu_s$ and 
 $p(\omega \given y, \theta, \mu_s) \propto N(\omega \given 0, \Omega) \times N(y \given 1_T\mu_s + \omega, \sigma^2I_T) \propto N(M_\omega m_\omega, M_\omega)$, where $m_\omega = \frac{1}{\sigma^2}(y - 1_T\mu_s)$ and $M_\omega = (\Omega^{-1} + \frac{1}{\sigma^2}I_T)^{-1}$. Thus, drawing one $\omega \sim N(M_\omega m_\omega, M_\omega)$ for each posterior sample of $\{\theta, \mu_s, y_{ns}\}$ will result in a set of posterior samples of $\omega$.

\textbf{Model 3. Two-Stage + Spatial} \\
The spatial model in (\ref{eq:sp}) can be rewritten using (\ref{eq:general_regression}) by letting $\theta = [\phi, \gamma^2, \delta^2, \tau^2, \sigma_1^2, \dots, \sigma_N^2]^\T$, with $p(\theta) = p(\phi) \times IG(\gamma^2\given a_\gamma, b_\gamma) \times IG(\delta^2 \given a_\delta, b_\delta) \times IG(\tau^2 \given a_\tau, b_\tau) \times \prod_{i=1}^NIG(\sigma_i^2 \given a_{\sigma_i}, b_{\sigma_i})$, $V_\nu = \gamma^2$, $V_\beta = \delta^2I_N$, $A = 1_N$, $V_s = \Omega_s + V_s^{(\sigma)}$, and $\beta = \mu$. After posterior samples of $\{\theta, \mu, y_{ns}\}$ are drawn as described in (\ref{eq:bayes_predict}), posterior samples of the spatial effects can be recovered by sampling one $\omega \sim N(M_{\omega2}m_{\omega2},M_{\omega2})$ for each posterior sample of $\{\theta, \mu, y_{ns}\}$, where $m_{\omega2} = V^{(\sigma) -1}(y - X\mu)$ and $M_{\omega2} = (\Omega^{-1} + V^{(\sigma) -1})^{-1}$.

\textbf{Model 4. Regional Spatial} \\
To rewrite the region-specific spatial model given using (\ref{eq:general_regression}), let $V_\nu = \gamma^2$, $V_\beta = \delta^2I_N$, $A = 1_N$, $V_s = \Omega_{s*} + V_s^{(\sigma)}$, and $\beta = \mu$. Also take $\theta = [\phi_1, \dots, \phi_N, \gamma^2, \delta^2, \tau_1^2, \dots, \tau_N^2, \sigma_1^2, \dots, \sigma_N^2]^\T$ with   $p(\theta) = \prod_{i=1}^Np(\phi_i) \times IG(\gamma^2\given a_\gamma, b_\gamma) \times IG(\delta^2 \given a_\delta, b_\delta) \times \prod_{i=1}^N IG(\tau_i^2 \given a_{\tau_i}, b_{\tau_i}) \times \prod_{i=1}^NIG(\sigma_i^2 \given a_{\sigma_i}, b_{\sigma_i})$. Similar to model 1, as not all locations are sampled, informative priors must be placed on the $\phi_{n+1}, \dots, \phi_N$ spatial decay parameters. Additionally, to recover posterior samples of $\omega$, sample one $\omega \sim N(M_{\omega3}m_{\omega3},M_{\omega3})$ for each posterior sample of $\{\theta, \mu, y_{ns}\}$ drawn using (\ref{eq:bayes_predict}), where $m_{\omega3} = V^{(\sigma) -1}(y - X\mu)$ and $M_{\omega3} = (\Omega_*^{-1} + V^{(\sigma) -1})^{-1}$.

To achieve computation efficiency, redefine $y = [y_1^\T, y_1^{*\T}, \dots, y_n^\T, y_n^{*\T},y_{n+1}^{*\T},\dots y_N^{*\T}]^\T$ so that the outcome is organized by region and then $\Omega_*$ becomes a $T \times T$ block diagonal matrix composed of $N$ blocks. This allows us to instead invert $N$ covariance matrices of size $M_1 \times M_1, \dots, M_N \times M_N$, rather than one $T \times T$ matrix, in the estimation of $\omega$. 
\subsection{Exact Monte Carlo Estimation} \label{sec:exact_models}
If we are able to provide reasonable fixed values of the parameters, (\ref{eq:bayes_regression}) can be simplified into a conjugate Bayesian linear model resembling:
\begin{equation}\label{eq:conjugate_bayes}
IG(\delta^2 \given a, b) \times N(\nu \given 0, \delta^2 \tilde{V}_\nu) \times N(\beta\given A \nu, \delta^2\tilde{V}_\beta) \times N(y_s \given X\beta, \delta^2\tilde{V}_s) \;.
\end{equation}
For a model such as this, the components $a$, $b$, $\tilde{V}_\nu$, $\tilde{V}_\beta$, and $\tilde{V}_s$ are fixed, reducing the model to three unknown parameters, $\delta^2$, $\nu$, and $\beta$. Thus, we can avoid MCMC and sample from the joint posterior $p(\delta^2, \nu, \beta \given y_s)$ using the following steps. First sample $\delta^2$ from $p(\delta^2 \given y-S) = IG(a^*, b^*)$ and then for each $\delta^2$ drawn, draw a corresponding $\nu$ from $N(M_\nu m_\nu, \delta^2M_\nu)$. Next, for each pair of $\{\delta^2, \nu\}$, draw $\beta$ from $N(M_\beta m_\beta, \delta^2M_\beta)$ (see the Supporting Information for details). As an example, we recast each model presented in Section~\ref{sec:mcmc} in the form of (\ref{eq:conjugate_bayes}) and derive the posterior conditional distributions for model 1 and model 2, details of which are provided in the Supporting Information.

\textbf{Model 1. Two-Stage} \\
To create a conjugate Bayesian model such as (\ref{eq:conjugate_bayes}) from the non-spatial model, define $A= 1_N$, $\tilde{V}_\nu = \tilde{\gamma}^2 = \frac{\gamma^2}{\delta^2}$, $\tilde{V}_\beta = I_N$,  and $\tilde{V}_s = \tilde{V}_s^{(\sigma)} = [\oplus_{i=1}^n \frac{\sigma_i^2}{\delta^2}I_{m_i}]$. Noting that $p(\nu \given y_s)  \propto N(\nu \given 0, \delta^2 \tilde{\gamma}^2) \times N(y_s \given X_{s1} 1_n \nu, \delta^2 [X_{s1} X_{s1}^\T + \tilde{V}_s^{(\sigma)} ])$ a little algebra reveals 
\begin{equation}\label{eq:nu_exact}
\nu \given y_s, \delta^2 \sim N(\nu\given c , \delta^2 d) \;,
\end{equation}
where $c = \frac{\sum_{i=1}^n  \lambda_i \bar{y}_i}{\frac{1}{\tilde{\gamma}^2} + \sum_{i=1}^n \lambda_i}$ and 
$d = \Big[\frac{1}{\tilde{\gamma}^2} + \sum_{i=1}^n \lambda_i\Big]^{-1}$.
The mean of the posterior distribution, $c$, is the weighted average of the sampled group means, where each mean is weighted by a function of each group's element-wise variance. Integrating out $\nu$ and $\mu$ from $p(\delta^2,\nu,\mu\given y_s)$ yields $p(\delta^2\given y_s)$, which is:
\begin{equation}\label{eq:delta_exact}
\delta^2\given y_s \sim IG\Bigg(a+\frac{k}{2}, b + \frac{1}{2}\Big[y_s^\T(X_{s1}X_{s1}^\T + \tilde{V}_s^{(\sigma)})^{-1}y_s + \frac{c^2}{d}\Big] \Bigg) \; .
\end{equation}
Taking the limits of $c$ and $d$ as $\tilde{\gamma}^2 \to\infty$ (e.g. $\gamma^2 \to \infty$) we recover the findings of \cite{Scott:1969fp}, who assigned $p(\nu) \propto 1$:
\[
\lim_{\tilde{\gamma}^2\to\infty} c = \frac{\sum_{i=1}^n  \lambda_i \bar{y}_i}{\sum_{i=1}^n \lambda_i } \; \text{and} \;
\lim_{\tilde{\gamma}^2\to\infty} \delta^2 d =  \frac{\delta^2}{\sum_{i=1}^n \lambda_i } \;.
\]
As $p(\mu_s \given y_s, \delta^2, \nu) \propto N(\mu_s \given \nu 1_n, \delta^2 I_n) \times N(y_s \given X_{s1} \mu_s, \delta^2 \tilde{V}_s^{(\sigma)})$ we have that: 
\begin{equation}\label{eq:mu_exact}
\mu_s \given y_s,\nu, \delta^2 \sim N(\mu_s \given c_*, \delta^2 d_*) \;,
\end{equation}
where $c_* = \begin{bmatrix} (1 - \lambda_1) \nu + \lambda_1\bar{y}_1 \\ \vdots \\ (1 - \lambda_n) \nu + \lambda_n\bar{y}_n\end{bmatrix}$ and
$d_* =  \begin{bmatrix} \oplus_{i=1}^n (1 - \lambda_i) \end{bmatrix}$. The posterior mean is appealing for interpretation, as its $i$-th element is the weighted average of the $i$-th group's sample mean and the superpopulation mean estimate. Finally, note that $\mu_{ns} \given y_s, \nu, \delta^2 \sim N(\mu_{ns} \given \nu, \delta^2 I_{N-n})$, as $y_s$ provides no information pertaining to the nonsampled groups.

\textbf{Model 2. Spatial} \\
The spatial model can be recast as (\ref{eq:conjugate_bayes}) by defining $\tilde{V}_\nu^{-1} = 0$ and $\tilde{V}_s = \tilde{\Omega}_s = \frac{1}{\delta^2} \Omega_s + I_k$, where $\tilde{V}_\beta$ is fixed to $1$. Defining $V_{\tilde{\Omega}_s} = (1+ 1_k^\T\tilde{\Omega}_s^{-1}1_k)^{-1}$, the posterior conditionals are:
\begin{equation}\label{eq:tau2_exact}
\delta^2 \given y_s \sim IG\left[a + \frac{k}{2}, b + \frac{1}{2}y_s^\T\Big(\tilde{\Omega}_s^{-1} - \tilde{\Omega}_s^{-1}1_kV_{\tilde{\Omega}_s} 1_k^\T\tilde{\Omega}_s^{-1} \Big)y_s\right] \text{and}
\end{equation}
\begin{equation}\label{eq:mu2_exact}
\mu_s \given y_s, \delta^2 \sim N\Bigg[V_{\tilde{\Omega}_s} 1_k^\T\tilde{\Omega}_s^{-1}y_s, \delta^2V_{\tilde{\Omega}_s} \Bigg] \;.
\end{equation}

\textbf{Model 3. Two-Stage + Spatial} \\
The form of (\ref{eq:conjugate_bayes}) is achieved by defining  $\tilde{V}_\nu = \tilde{\gamma}^2$, $\tilde{V}_\beta = I_N$, $A = 1_N$, and $\tilde{V}_s = \frac{1}{\delta^2} \Omega_s + \tilde{V}_s^{(\sigma)} $.

\textbf{Model 4. Regional Spatial} \\
The form of (\ref{eq:conjugate_bayes}) is achieved by defining $\tilde{V}_\nu = \tilde{\gamma}^2$, $\tilde{V}_\beta = I_N$, $A= 1_N$, and $\tilde{V}_s = \frac{1}{\delta^2} \Omega_{s*} + \tilde{V}_s^{(\sigma)} $.
 
\subsection{Model Comparison and Assessment} \label{sec:model_assessment}
Model fit was evaluated in two ways. In general, consider a sample of size $k$ drawn from a population of size $T$ with outcome $y = [y_s ^\T\colon y_{ns}^\T]^\T$. Without loss of generality, say $y_h \in y_s$ if $h = 1, \dots, k$ and $y_h \in y_{ns}$ if $h = k + 1, \dots, T$. First we evaluate the predictive accuracy of the models using the Watanabe-Akaike Information Criteria (WAIC), which is expressed as $WAIC = -2\widehat{elpd} = -2(\widehat{lpd} + \hat{p}_{WAIC})$ in \cite{Vehtari:2017wy}, where $\widehat{elpd}$ is the estimated expected log pointwise predictive density and is multiplied by $-2$ to be on the deviance scale.
To calculate this, at each iteration, $l = 1, \dots, L$, $p(y_h \given \Theta^{(l)})$ is computed; the likelihood of each observed value conditional on that iteration's parameters. The estimated log pointwise predictive density is the sum of the log average likelihood for each observation, $\widehat{lpd} = \sum_{h=1}^k \log\Big[\frac{1}{L} \sum_{l=1}^L p(y_h \given \Theta^{(l)})\Big]$. The sample variance of the log-likelihood for each observation is $s_{lp(y_h)}^2 = \frac{1}{L-1} \sum_{l=1}^L\Big[\log(p(y_h \given \Theta^{(l)})) - \frac{1}{L} \sum_{l=1}^L \log(p(y_h \given \Theta^{(l)}))\Big]^2$ and the estimated effective number of parameters is the sum of these variances: $\hat{p}_{WAIC} = \sum_{h=1}^k s_{lp(y_h)}^2$. To calculate the standard error of the WAIC, rewrite $-2(\widehat{lpd} + \hat{p}_{WAIC}) = -2 \sum_{h=1}^k \widehat{elpd}_h = \sum_{h=1}^k \Big\{\log\Big[\frac{1}{L} \sum_{l=1}^L p(y_h \given \Theta^{(l)})\Big] + s_{lp(y_h)}^2 \Big\}$. Under the assumption that each $\widehat{elpd}_h$ is independent, the sample variance of each individual $\widehat{elpd}_h$ is $s_{elpd,ind}^2 = \frac{1}{N-1} \sum_{h=1}^k\Big[\widehat{elpd}_h - \frac{1}{k}\sum_{h=1}^k \widehat{elpd}_h \Big]$. Then $SE(WAIC) = \sqrt{Var(-2 \sum_{h=1}^k \widehat{elpd}_h)} = 2 \sqrt{nVar(\widehat{elpd}_h)} = 2 s_{elpd,ind}\sqrt{n}$.

Second, for simulated data the true values $y= [y_s : y_{ns}]^\T$ are known and so we compare these with replicated datasets, $y_{rep}^{(l)} = [y_{rep,1}^{(l)} \dots y_{rep,k}^{(l)}]^\T$, generated from the pointwise posterior predictive distribution at each iteration $l$. These are used to formulate the goodness of fit measurement $D = G + P$ described in \cite{Gelfand:1998wj}, composed of an error sum of squares term and a penalty term for large predictive variances. For $L$ iterations, $G = \sum_{h=1}^k (y_h -\text{E}[y_{rep,h} \given y_s])^2$ and $P = \sum_{h=1}^k var(y_{rep,h}\given y_s)$. We approximate $\text{E}[y_{rep,h} \given y_s] \approx  \frac{1}{L}\sum_{l=1}^L y_{rep,h}^{(l)}$ and $var(y_{rep,h}\given y_s)  \approx \frac{1}{L-1} \sum_{l=1}^L (y_{rep,h}^{(l)} - \frac{1}{L}\sum_{l=1}^L y_{rep,h}^{(l)})^2$. For non-simulated datasets, where $y_{ns}$ is unknown, $D$ can still be calculated by restricting the replicate datasets to the observed units, $y_s$, e.g. $y_{rep}^{(l)} = [y_{rep,1}^{(l)} \dots y_{rep,k}^{(l)}]^\T$, at the $l$-th iteration. 

\section{Simulation}\label{sec:sim}
\subsection{Data Generation}
To simulate spatial correlation and allow for two-stage random sampling, a unit square was divided into 100 equally sized square regions and 2,500 locations were randomly drawn from the unit square. Data was simulated from the intercept-only spatial model described in Model 2 with $\mu = 2$. A distance matrix for all points was constructed and used to create a covariance matrix that reflects an exponential covariance function described in Section~\ref{sec: bayesian_spatial_process_modeling}, where $\phi$ was assigned a value of 10, reflecting an effective spatial range of 3/10. The spatial variance, $\tau^2$, was fixed at 9, while the non-spatial variance, $\sigma^2$, was set to 4. After a dataset was generated, a cluster random sampling scheme was implemented. 25 regions were randomly selected and then in each cluster, a random number of individuals were selected (the minimum and maximum percent of those selected from a region was set to be 20\% and 90\%, respectively). 20 datasets containing information of both the sampled and nonsampled units were generated in this way. To examine Models 1 and 4 for larger dataset, this process was then repeated with the same parameters to generate 20 datasets with 8,100 locations from 324 regions, where 81 regions were randomly sampled. All data generation and analyses were performed using R version 3.5.1 \citealt{Rlang}.

\subsection{Exact Monte Carlo Simulation} \label{sim.exact}

To perform the two-stage procedure using the conditional distributions and methods described in Section~\ref{sec:exact_models}, sample means, $\hat{\mu}_i$, and sample variances, $\hat{\sigma}_i^2$, were calculated from each observed cluster, $i = 1, \dots, n$. The variance matrix of the sampled units was fixed to be $V_s^{(\sigma)} = \begin{bmatrix} \oplus_{i = 1}^n \frac{\hat{\sigma}_i^2}{\text{Var}(\hat{\mu})} I_{m_i}\end{bmatrix}$, where $\text{Var}(\hat{\mu})$ represents the sample variance of the observed sample means. Similarly, the variance matrix of the nonsampled units was fixed at $V_{ns}= \begin{bmatrix} \oplus_{i = 1}^N \frac{\tilde{\sigma}_i^2}{\text{Var}(\hat{\mu})} I_{M_i - m_i}\end{bmatrix}$, where $\tilde{\sigma}_i^2 = \hat{\sigma}_i^2$ if $i \in \{1, \dots, n\}$ and $\tilde{\sigma}_i^2 = \frac{1}{n}\sum_{i=1}^n \hat{\sigma}_i^2$ if $i \in \{n+1, \dots, N\}$. The value of $\gamma^2$ was fixed to be half of the value of $\delta^2$, reflecting the belief that there was less variability in the population mean than between group means. The prior distribution for $\delta^2$ was assigned to be $IG(3,5)$. Sampling from the posterior was performed using the conditional distributions and methods described in Section~\ref{sec:exact_models}. At each iteration $g$, the population mean estimate for that iteration was then calculated as $\bar{y}^{(g)} = \frac{1}{T}\Big(\sum_{i = 1}^n\sum_{j=1}^{m_i} y_{ij} + \sum_{i = 1}^N\sum_{j=m_i + 1}^{M_i} y_{ij}^{(g)}\Big)$. Details of this iterative procedure can be found in the supplementary materials. 

To perform the spatial random effect procedure, $\phi$ was set to its true value of 10 and the ratio of $\delta^2/\tau^2$ to its true value of $4/9$. The posterior conditionals of $\delta^2 \given y_s$ and $\mu_s \given y_s, \delta^2$, (\ref{eq:tau2_exact}) and (\ref{eq:mu2_exact}) respectively, were sampled as outlined in Section~\ref{sec: bayesian_spatial_process_modeling}. This sampling and the prediction of $y_{ns}$ was performed using commands from the \textit{spBayes} R package \citep[]{Finley:2015tb, Finley:2007fb}. The population mean estimate was calculated using the technique described in the non-spatial sampling case above.

\begin{figure}[ht]
\centerline{%
\includegraphics[width= \textwidth]{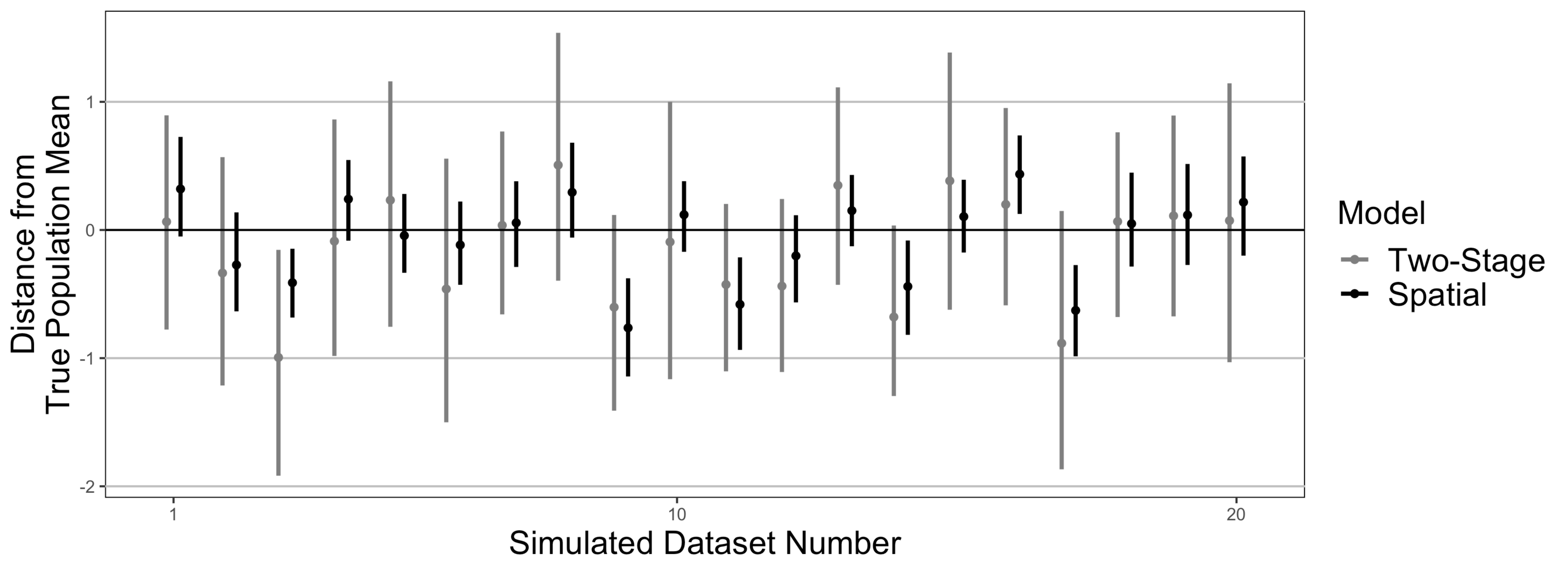}}
\caption{Centered Population Mean Estimates from 2 Exact Models with 95\% CI}
\label{fig:sim1}
\end{figure}

Figure~\ref{fig:sim1} plots population average-centered mean estimates and 95\% credible intervals from both methods applied to the twenty simulated datasets. While the spatial cases consistently have a smaller credible interval, their point estimates are similar to the two-stage case. However, as the ratio of spatial and non-spatial variance is fixed for this method, this may result in a reduction in the overall variance of the population mean. Posterior mean estimates and their associated 95\% credible intervals of the superpopulation parameters and finite population mean, $\bar{y}$, from the first generated dataset are given in Table~\ref{tab:sim1}, along with the WAIC, its standard error, and $D$ values. While both models have similar estimates for $\nu$, the two-stage model overestimates the non-spatial variance. This is expected, as we know that there is additional variance due to spatial correlation that is not being accounted for otherwise in the model. Similarly, both measures of goodness of fit prefer the spatial model.

\begin{table}[ht]
\caption{Comparison of Parameter Estimation and Model Fit in Two Exact Models}
\centering
\begin{tabular}{lll}
  \hline
 & Two-Stage & Spatial \\ 
  \hline
  $\nu$ (2) & 2.60 (1.55, 3.58) &2.79 (1.36, 4.23) \\ 
  $\delta^2$ (4) & 6.36 (5.47, 7.45) & 3.84 (3.35, 4.43) \\
  $\tau^2$ (9) & --- & 8.65 (7.53, 9.96) \\
  $\bar{y}$ (2.60) & 2.66 (7.53, 9.96) & 2.92 (7.53, 9.96) \\
  WAIC & 1803.83 (25.52) & 1686.7 (27.6)\\ 
  D = G + P & 66100.83=34203.26+31897.56 & 45335.54=22875.45+22460.08 \\ 
   \hline
\end{tabular}
\bigskip
\label{tab:sim1}
\end{table}

\subsection{Markov Chain Monte Carlo Simulation} \label{sec:sim_mcmc}
To explore these findings further, we implemented the four models described in Section~\ref{sec:models} using the JAGS software in R on the same generated datasets. Models were run for 650 iterations with 50 burn-in, as examination of individual trace plots suggested sufficient mixing and convergence of the non-spatial parameters. At each iteration $g$, estimates of the nonsampled units were drawn and estimates for the population mean, $\bar{y}^{(g)} = \frac{1}{T}\Big(\sum_{i = 1}^n\sum_{j=1}^{m_i} y_{ij} + \sum_{i = 1}^N\sum_{j=m_i + 1}^{M_i} y_{ij}^{(g)}\Big)$ were calculated. All variance parameters ($\sigma^2$, $\tau^2$, and $\delta^2$, as well as site-specific variances such as $\sigma_i^2$ and $\tau_i^2$) were given an inverse-gamma prior with shape 2 and scale 10, reflecting a weakly-informative prior distribution with mean $10$. There is a substantial literature in theoretical spatial statistics regarding the identifiability, or lack there of, of the spatial process parameters, hence, non-informative or completely flat priors are excluded from consideration. The prior families and specifics we use are fairly customary in spatial modeling. They exploit some information about the spatial domain and the extent of spatial association that can be detected from finite samples using variograms. For example, in practice, given a real data set, we would pass the data through an exploratory analysis tool, such as a variogram, glean some information about the spatial variance component and the measurement error component, and use the weakly informative centered inverse-gamma priors to reflect these values. In addition, $\nu$ was given a flat prior to not inform the estimation of the mean and all $\phi$ parameters were given Uniform(5,15) priors to allow the spatial range to vary from 0.2 (3/15) to 0.6 (3/5). While our priors were chosen to be weakly-informative to be conservative in estimation, more informative priors could easily be added in a data analysis if additional information regarding the parameters was known. MCMC sampling was performed using the computer program JAGS \citep{Plummer:2017vp} in R.

When assessing model fit in the first realization of the data with WAIC, the spatial model performed slightly worse than the rest of the models with a value of 1,912.70 (SE = 26.36). This may be due to the additional variation which comes from varying the spatial range parameter. This was followed closely by the two-stage model with 1,870.26 (35.00) , which was outperformed by both the regional spatial model with 1,202.08 (38.99) and the two-stage + spatial model with 455.67 (17.03). It is interesting that while the data was generated by the spatial model and sampled by a two-stage framework, neither of these models perform better than the two models which take both the spatial correlation and study design into account. Additional comparisons of estimates from this first realization can be found in the supplementary materials.

\begin{figure}[ht]
\centerline{%
\includegraphics[width = \textwidth]{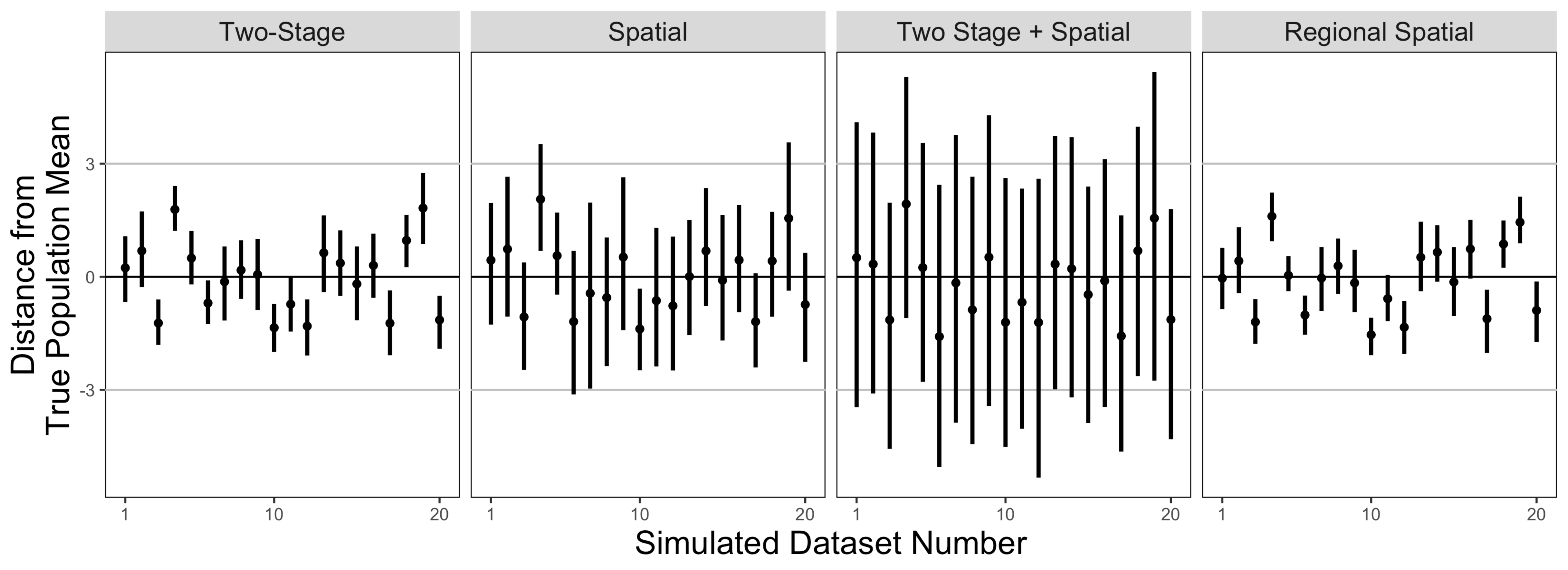}}
\caption{Centered Population Mean Estimates from 4 MCMC Models with 95\% CI}
\label{fig:sim2}
\end{figure}

Figure~\ref{fig:sim2} shows the models' posterior mean estimates of the finite population mean, which are centered at the true population mean and presented with 95\% credible intervals for the 20 simulated datasets. While point estimates remain similar across models, the best fitting model, Model 3, has the widest credible intervals for the population mean. Accounting for only regional effects results in tight credible intervals in Models 1 and 4, which are narrow compared to Model 2, which fails to take into account region specific variability. Similar results were found when applying Models 1 and 4 to the larger simulated datasets and are provided in the supplementary materials.

\section{Data Analysis: Nitrate in Central California Groundwater}\label{sec:data_analysis}
In this section, we provide an analysis of groundwater nitrate content of the Tulare Lake Basin (TLB) in Central California from the California Ambient Spatio-Temporal Information on Nitrate in Groundwater (CASTING) Database, which is described in \cite{Harter:2017vb} and \cite{Boyler:2012wv} and is available as the University of California, Davis, nitrate data in the data repository of the \cite{GAMA}. Interest lies in identifying regions in which ground water nitrate levels exceed 45 mg/L, which is the maximum contaminant level established by the EPA \cite{Boyler:2012wv}. At high levels, infants and pregnant women are more susceptible to nitrate poisoning, which makes it more difficult for oxygen to be distributed to body and can be fatal to infants less than six months old. Besides human sources such as sewage disposal, many sources of nitrate are agricultural, such as fertilizer for crops and animal waste \citep{Harter:2012va}. Because of this, regions with high agricultural activity, such as the TLB, have experienced rising levels of nitrate over the past few decades. As groundwater, and therefore nitrate levels in groundwater, can be assumed to be present at all areas of the Central Valley, we can assume that water samples taken from wells come from a spatial field. Therefore, given a sample of readings from various wells, our primary goal is to estimate of the finite population average of all known wells, which represents an overall measure of water-health. Additionally, plots of posterior predictive distribution may be useful in identifying high-risk regions which exceed the maximum contaminant level. 

The CASTING Database is an extensive collection of nitrate readings from the TLB and Salinas Valley collected by multiple agencies, over 70\% of which were collected between 2000 and 2011. Of these, most wells had repeated measurements taken over the time. As the Salinas Valley and the TLB are geographically separate regions of California, only the TLB was included. In order to avoid associations over time, the data was restricted to a single year. The year 2009 was selected as variogram plots suggested nitrate levels followed a roughly exponential distribution. While directional variograms suggested that the measurements may be anisotropic, we continued with the methods presented above, recognizing that a model accounting for directional spatial dependence may provide a better fit to the data. 

While the data roughly covers the TLB, we construct a likely sampling scenario in which the TLB is separated into distinct geographic regions and due constraints (perhaps time or financial), a random subset of these regions are sampled. In our scenario, zip codes were used as it is common to collect such geographic information in large scale health surveys, but many alternatives such as cities or grid-based approach could have also been used. A map of California zip codes tabulation areas obtained from the \textit{tigris} R package \citep{Walker:2018ti} was overlaid on the approximate geographic locations of each of the sampled wells, effectively assigning each well to one specific region, defined by a zip code. 1) Only the most recent observation was taken from each well so that each well was only represented once. 2) If unique wells had the same geographic coordinates, one was chosen at random to be removed. 3) Sparse zip codes with less than 10 wells were excluded to ensure that each selected zip code would have a large sample size and to avoid overfitting when modeling. These restrictions resulted in a dataset with 6,117 unique wells among 63 zip codes. Nitrate level had a mean of 37.9 mg/L, standard deviation of 52.3 mg/L, and ranged from 0.0 to 903.1 mg/L.

In order to recreate a cluster sampling scenario, 21 of the zip codes were randomly chosen and 50-90\% of the population in that zip code was randomly sampled. This resulted in an observed sample size of 489 with a mean nitrate level of 34.2 mg/L and a standard deviation of 40.0 mg/L. The nitrate level ranged from 0.0 to 269.6. A plot of these sampled and non-sampled zip codes is available in the supplementary materials. Using this sampled data, all four models in Section~\ref{sec:sim_mcmc} were implemented and the results are shown in Table ~\ref{tab:da}. In all models, $\nu$ was given a flat prior. For model 1, the regional variance parameters were given an inverse-gamma prior with shape 2 and scale 1600 to reflect the sample variance. For the spatial models, a variogram was fit and spatial variances were given an inverse-gamma prior with shape 2 and scale 1800. Similarly, non-spatial variance terms were assigned an inverse-gamma prior with shape 2 and scale 1000. The variance of regional means was assigned an inverse-gamma prior with shape 2 and scale 10 to allow for small, localized deviations. All $\phi$ parameters were given Uniform(0.01,5) priors to reflect a spatial range varying between 0.6 km (3/5) and 300 km (3/0.01). MCMC sampling was performed using JAGS \citep{Plummer:2017vp} in R \citep{Rlang}.

With respect to the estimate of the true mean nitrate level, only the intercept-only spatial model contained the true mean value within its 95\% credible intervals. However, as evidenced by the larger mean, standard deviation, and range in the complete dataset, it appears that the sampled units did not capture some of the larger outliers, so it is unsurprising that the estimates of the population mean are lower than the truth. Comparing WAIC, we see results similar to those found in Section~\ref{sec:sim_mcmc}. The spatial models which accounted for regional means had lower WAIC values than the two-stage model, which is evidence that this data is spatially correlated. However, the intercept-only spatial model did not fit the data as well as the two-stage model, which may be due to ignoring the study design. Additionally, the two-stage + spatial model again fits the model the best. 

\begin{figure}[ht]
\centerline{
\includegraphics[width = \textwidth]{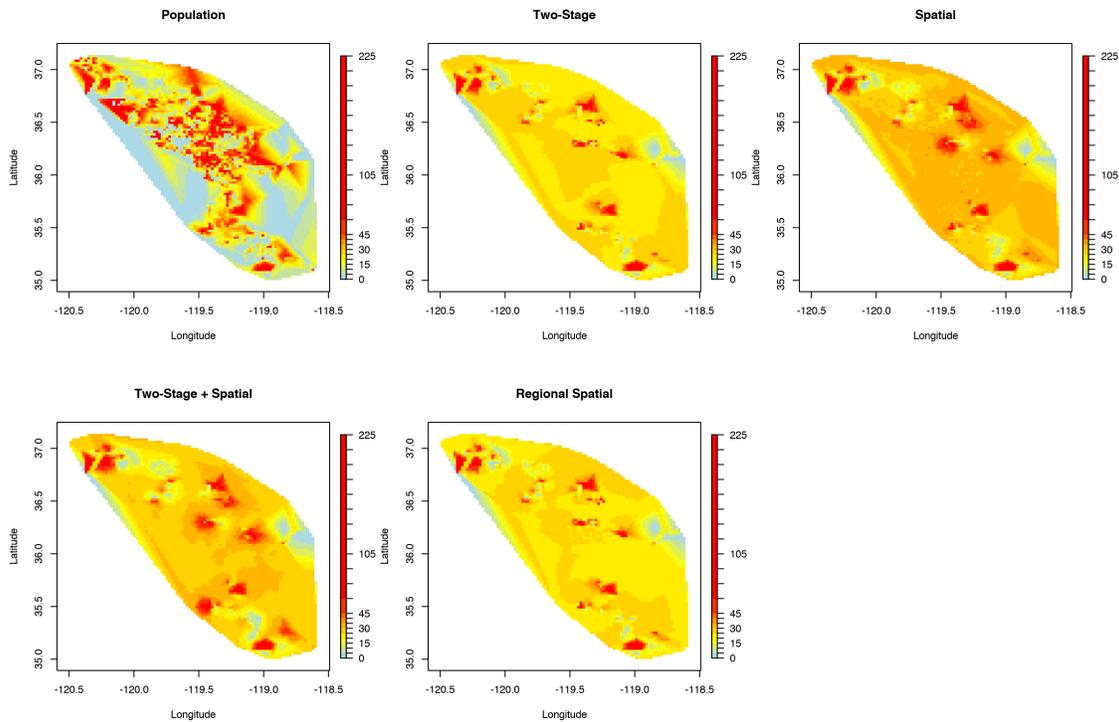}}
\caption{Interpolated surfaces using the population (truth) and posterior predictive samples from the four models.}
\label{fig:da}
\end{figure}

Figure~\ref{fig:da} shows the interpolated population surface from the complete sample and the interpolated surface from posterior predictive samples. While there are common regions at high risk (nitrate level greater than 45 mg/L) in all the posterior predictive maps, the spatial and two-stage + spatial maps predict larger regions. Also seen in Table~\ref{tab:da}, it is clear that Model 3 estimates a population mean that is larger than Models 1 and 4, but smaller than Model 3. Spatial residual plots for Models 2, 3, and 4 are provided in the supplementary materials.

\begin{table}[ht]
\caption{Results of Data Analysis}
\centering
\begin{tabular}{lll}
  \hline
 Model & FP Mean (95\% CI) & WAIC (SE) \\ 
  \hline
  1. Two-Stage & 26.6 (19.2, 35.1) & 4701.6 (78.6) \\ 
  2. Spatial & 32.6 (27.2, 38.7)  & 4858.4 (78.0) \\ 
  3. Two-Stage + Spatial & 31.2 (24.2, 37.3) & 2697.0 (150.5) \\ 
  4. Regional Spatial & 26.2 (18.7, 34.4) & 4536.6 (93.1) \\ 
   \hline
\end{tabular}
\bigskip
\label{tab:da}
\end{table}

\section{Discussion}\label{sec:Discussion}
This paper examines the implications of performing two-stage random sampling on point-referenced data which exists in a spatial field. While \cite{Scott:1969fp} and \cite{Malec:1985tr} provided a Bayesian model-based framework to account for such a study design, we have demonstrated that an analysis ignoring the underlying spatial correlation between locations or sampling design may lead to spurious inference and poorer model fit. 

This work is a first step in developing an overarching framework for Bayesian finite population sampling from spatial process based populations. In our two-stage case, additional work may be done to further improve this model. For instance, CAR priors could be placed on regional parameters such as the $\mu_i$'s, the regional means, to induce additional spatial correlation in the model. Also, many of the models presented account for regional differences in variance but if other sources of heteroscedasticity are suspected, new approaches (such as \citealt{Zangeneh:2015ab}) may be needed to account for this. While an exponential covariance function was employed in the analyses in this paper, other spatial covariance functions could be used to create similar simulations and data analyses, as well as account for anisotropy.

Future work is needed to establish a more general framework that can account for more sophisticated sampling designs in a spatial context. The sampling designs presented in this paper are said to be ignorable (\citealt{Rubin:1976we}; \citealt{Sugden:1984tw}), which allows us to perform inference on the superpopulation parameters while ignoring the inclusion probability distribution. However, designs in which the data cannot be assumed to be missing at random or where parameters define both the outcome and inclusion distributions are referred to as nonignorable and must account for the inclusion probability distribution. One example of this in the spatial context is preferential sampling (\citealt{Diggle:2010ir}, \citealt{Gelfand:2012ix}), in which the measurement values and sampling strategy are assumed to stem from the same spatial process. While \cite{Pati:2011gd} have analyzed such data using Bayesian hierarchical models, an overall framework is needed to account for this and other non-ignorable design types.

Additionally, the implications of study design on finite population estimates when sampling from a spatially correlation population over time are unknown. In order to better understand this, these Bayesian models must first be extended to account for both study design and spatio-temporal associations.

Finally, while this paper provided a scaleable model which can account for study design and spatial correlation in massive survey data by assuming regional independence, further work should be done to incorporate recent strategies in modeling large spatial data \citep{Heaton:2018tp} when analyzing survey data with spatial correlations, such as nearest neighbor processes \citep{Datta:2016xj}, covariance tapering \citep{Furrer:2006pn}, and metakriging \citep{Guhaniyogi:2018ab}. Finite population models would particularly benefit from such techniques, as computation increases as a function of the population total, $T$, rather than the sample size, $k$. 

\section*{Supporting Information}
The work of the first and second authors were supported, in part, by the Division of Information and Intelligent Systems of the National Science Foundation under Grant 1562303, the Division of Mathematical Sciences of the National Science Foundation under Grant 1916349, and the National Institute of Environmental Health Sciences of the National Institutes of Health under Grants 1R01ES027027 and R01ES030210-01.

\bibliographystyle{ba}  
\bibliography{bib_bfps} 

\newpage
\appendix

\section*{Appendix A}

This section first provides the derivation of the empirical Bayesian estimators presented in Section~\ref{sec:exact_models} and then the finite population estimates presented in Sections~\ref{sec: bayesian_multi_stage_sampling} and \ref{sec: bayesian_spatial_process_modeling}. 

The values $a^*$, $b^*$, $M_\nu$, $m_\nu$, $M_\beta$, and $m_\beta$ for the general case (\ref{eq:conjugate_bayes}) in Section~\ref{sec:exact_models} are presented below. 
\begin{align*}
a^* &= a + \frac{n}{2} \\
b^* &= b + \frac{1}{2}\big(y_s^\T y_s - y_s^\T \tilde{V}_s^{-1}X_s^\T [(\tilde{V}_\beta + A\tilde{V}_\nu A^\T)^{-1} + X_s^\T \tilde{V}_s^{-1}X_s ]^{-1}X_s^\T \tilde{V}_s^{-1}y_s\big)\\
M_\nu &= \big[\tilde{V}_\nu^{-1} - A^\T \tilde{V}_\beta^{-1}(\tilde{V}_\beta^{-1} + A^\T X_s^\T\tilde{V}_s^{-1}X_sA)^{-1}\tilde{V}_\beta^{-1}A^\T\big]^{-1}\\
m_\nu &= A^\T \tilde{V}_\beta^{-1}(\tilde{V}_\beta^{-1} + A^\T X_s^\T\tilde{V}_s^{-1}X_sA)^{-1}A^\T X_s^\T \tilde{V}_s^{-1} y_s\\
M_\beta &= \big[ \tilde{V}_\beta^{-1} + A^\T X_s^\T \tilde{V}_s^{-1}X_sA\big]^{-1}\\
m_\beta &= \tilde{V}_\beta^{-1}A\nu + A^\T X_s^\T \tilde{V}_s^{-1}y_s
\end{align*}
The derivation of these values arise the conjugacy of the Normal and Inverse-Gamma distributions. We first derive (\ref{eq:delta_exact}) and (\ref{eq:mu_exact}). Take $\epsilon_s \sim N(0,\delta^2 \tilde{V}_s^{(\sigma)} )$, $\epsilon_{ns} \sim N(0,\delta^2 \tilde{V}_{ns}^{(\sigma)} )$, and $\nu \sim N(0,\delta^2 \tilde{\gamma}^{2})$, where $\tilde{\gamma}^2 = \frac{1}{\delta^2} \gamma^2$, $\tilde{V}_s^{(\sigma)}  = \frac{1}{\delta^2} V_s^{(\sigma)} = \begin{bmatrix} \oplus_{i = 1}^n \frac{\sigma_i^2}{\delta^2} I_{m_i}\end{bmatrix}$, and $\tilde{V}_{ns}^{(\sigma)}  = \frac{1}{\delta^2}V_{ns}^{(\sigma)} = \begin{bmatrix} \oplus_{i = 1}^n \frac{\sigma_i^2}{\delta^2} I_{(M_i-m_i)}\end{bmatrix}$. Since the elements of $y$ are independent conditional on $\mu$, $V_{s,ns} = 0$ and $V_{ns,s} = 0$. Define observed group means as $\bar{y}_i = \frac{1}{m_i}\sum_{j=1}^n y_{ij}$ and the ratio of variances as $\lambda_i = \delta^2/(\delta^2 + \sigma_i^2/m_i)$ if $i \in (1, \dots, n)$ and $\lambda_i = 0$ if $i \in (n+1, \dots, N)$. Also define the vector of observed group variances to be $\tilde{\sigma}^2 = [\sigma_1^2, \dots, \sigma_n^2]^\T$. Recall $\mu_s = \nu1_n + \epsilon_{\mu_s}; \epsilon_{\mu_s} \sim N(0,\delta^2I_n )$, then $y_s = X_{s1} \mu_s + \epsilon_s
= X_{s1} 1_n \nu + e_s^*$, where $e_s^* \sim N(0, \delta^2 [X_{s1} X_{s1}^\T + \tilde{V}_s^{(\sigma)} ])$. Then we have that $p(\nu \given y_s) \propto N(\nu \given 0, \delta^2 \tilde{\gamma}^2) \times N(y_s \given X_{s1} 1_n \nu, \delta^2 [X_{s1} X_{s1}^\T + \tilde{V}_s^{(\sigma)} ]) \propto N(\nu\given Bb , \delta^2 B)$, where
\begin{align*}
 b &= 1_n^\T X_{s1} ^\T ( X_{s1} X_{s1}^\T + \tilde{V}_s^{(\sigma)} )^{-1} y_s = \begin{bmatrix} 1_{m_1}^\T, \dots,  1_{m_n}^\T \end{bmatrix}
\begin{bmatrix} \oplus_{i=1}^n \Big(\frac{\delta^2}{\sigma^2_i}\Big) \Bigg(I_{m_i} - \frac{\frac{\delta^2}{\sigma^2_i}1_{m_i}1_{m_i}^\T}{1 + \frac{\delta^2}{\sigma^2_i}m_i} \Bigg) \end{bmatrix} y_s  \;\text{and}\\
B^{-1}&= \frac{1}{\tilde{\gamma}^2} + 1_n ^\T X_{s1} ^\T (X_{s1} X_{s1}^\T + \tilde{V}_s^{(\sigma)} )^{-1} X_{s1} 1_n 
=  \frac{1}{\tilde{\gamma}^2} +  \sum_{i=1}^n \Big(\frac{\delta^2}{\sigma^2_i}\Big) 1_{m_i}^\T \Bigg(I_{m_i} - \frac{\frac{\delta^2}{\sigma^2_i}1_{m_i}1_{m_i}^\T}{1 + \frac{\delta^2}{\sigma^2_i}m_i}\Bigg)1_{m_i} \;.
\end{align*}
Therefore, $B = \Big[\frac{1}{\tilde{\gamma}^2} + \sum_{i=1}^n \lambda_i\Big]^{-1}$ and 
$Bb = \frac{\sum_{i=1}^n  \lambda_i \bar{y}_i}{\frac{1}{\tilde{\gamma}^2} + \sum_{i=1}^n \lambda_i}$. To solve $p(\delta^2\given y_s)$, split the posterior conditional distribution of the superpopulation parameters; $p(\delta^2,\nu\given y_s) = IG(\delta^2\given a_\delta^*,b_\delta^*) \times N(\nu \given Bb,\delta^2B)$, where $a_\delta^* = a_\delta + \frac{k}{2}$ and 
$b_\delta^* = b_\delta + \frac{1}{2}[y_s^\T(X_{s1} X_{s1}^\T + \tilde{V}_s^{(\sigma)} )^{-1}y_s + b^\T Bb]$. As $p(\mu_{ns} \given y_s, \nu, \delta^2) = p(\mu_{ns} \given \nu, \delta^2)$, $\mu_{ns}\given \nu, \delta^2 \sim N(1_{N-n}\nu, \delta^2I_{N-n})$. To solve $p(\mu_s \given y_s, \nu, \delta^2)$, note that 
$p(\mu_s \given y_s, \nu, \delta^2) \propto N(\mu_s \given 1_n\nu , \delta^2 I_n) \times N(y_s \given X_{s1} \mu_s, \delta^2 \tilde{V}_s^{(\sigma)} )\propto N(\mu_s \given B_*b_*, \delta^2 B_*)$, where 
\begin{align*}
b_* &= \nu 1_n + X_{s1}^\T (\tilde{V}_s^{(\sigma)})^{-1} y_s
=\begin{bmatrix} \nu + \frac{\delta^2}{\sigma_1^2}  m_1 \bar{y}_1,\; \cdots,\; \nu + \frac{\delta^2}{\sigma_n^2}  m_n \bar{y}_n\end{bmatrix} ^\T \;, \\
B_*^{-1} &= I_n + X_{s1}^\T (\tilde{V}_s^{(\sigma)})^{-1} X_{s1} 
=  \begin{bmatrix} \oplus_{i=1}^n \frac{\sigma_i^2 + \delta^2 m_i} {\sigma_i^2}\end{bmatrix} \;;\;
B_* = \begin{bmatrix} \oplus_{i=1}^n \frac{\sigma_i^2}{\sigma_i^2 + \delta^2 m_i}\end{bmatrix} 
= \begin{bmatrix}\oplus_{i=1}^n (1 - \lambda_i)\end{bmatrix} \; \text{, and}\\
B_*b_* &= \begin{bmatrix} \oplus_{i=1}^n (1 - \lambda_i) \end{bmatrix} 
\begin{bmatrix} \nu + \frac{\delta^2}{\sigma_1^2}  m_1 \bar{y}_1, \;\cdots ,\;\nu + \frac{\delta^2}{\sigma_n^2}  m_n \bar{y}_n\end{bmatrix} ^\T \\
&= \begin{bmatrix} (1 - \lambda_1) \nu + \lambda_1\bar{y}_1, \; \cdots ,\; (1 - \lambda_n) \nu + \lambda_n\bar{y}_n\end{bmatrix} ^\T \;.
\end{align*}

To derive (\ref{eq:tau2_exact}) and (\ref{eq:mu2_exact}) define $\tilde{V}_\nu^{-1} = 0$ and $\tilde{V}_s = \tilde{\Omega}_s = \frac{1}{\delta^2} \Omega_s + I_k$, and $\tilde{V}_{\mu_s} = 1$.
Note that $p(\mu_s \given y_s, \delta^2) \propto N(\mu_s \given 0, \delta^2) \times N(y_s \given 1_k \mu_s, \delta^2\tilde{\Omega}_s) \propto N(\mu_s \given B_{\mu_s}b_{\mu_s}, \delta^2B_{\mu_s})$, where $B_{\mu_s} = (1+ 1_k^\T\tilde{\Omega}_s^{-1}1_k)^{-1}$ and $b_{\mu_s} = 1_k^\T\tilde{\Omega}_s^{-1}y_s$. Splitting the posterior conditional distribution of the superpopulation parameters, $p(\delta^2,\mu_s \given y_s) = IG(\delta^2\given a_\delta^{**,}b_\delta^{**}) \times N(\mu_s \given B_{\mu_s}b_{\mu_s},\delta^2B_{\mu_s})$, where $a_\delta^{**} = a + \frac{k}{2}$ and $b_\delta^{**} = b + \frac{1}{2}y_s^\T\Big\{\tilde{\Omega}_s^{-1} - \tilde{\Omega}_s^{-1}1_k\Big(1 +1_k^\T\tilde{\Omega}_s^{-1} 1_k\Big)^{-1}1_k^\T\tilde{\Omega}_s^{-1} \Big\}y_s$.

We now continue by deriving the general cases presented in Section~\ref{sec: bayesian_multi_stage_sampling}.
As $\beta \given \nu \sim \text{N}(A\nu, V_\beta)$ and $\nu \sim \text{N}(0,\gamma^2)$, we have  that  $\beta \sim \text{N}(0, \gamma^2  AA^\T +  V_\beta)$. Note that $p(\beta \given y_s) \propto N(0, \gamma^2  AA^\T +  V_\beta) \times N(X_s \beta, V_s) \propto N(V_{\beta \given y_s} X_s^\T V_s^{-1} y_s, V_{\beta \given y_s})$, where $V_{\beta \given y_s} = [(\gamma^2  AA^\T  + V_\beta)^{-1} + X_s^\T V_s^{-1} X_s]^{-1}$. Defining  $B_V  = V_{ns,s} + (X_{ns} - V_{ns,s}V_s^{-1}X_s)V_{\beta\given y_s}X_s^\T$ and $Q = X_{ns} - V_{ns,s}V_s^{-1}X_s$, we have that:
\begin{align*}
\text{E}[\alpha^\T y \given y_s] &= \alpha_s^\T y_s + \alpha_{ns}^\T \text{E}[\text{E}[y_{ns} \given \beta, y_s] \given y_s] \\
&=  \alpha_s^\T y_s + \alpha_{ns}^\T \text{E}[X_{ns}\beta + V_{ns,s}V_s^{-1}(y_s - X_s\beta) \given y_s] \\
&=  \alpha_s^\T y_s +  \alpha_{ns}^\T V_{ns,s}V_s^{-1}y_s + QV_{\beta \given y_s}X_s^\T V_s^{-1} y_s  \\
&=  \{\alpha_s^\T  +  \alpha_{ns}^\T [V_{ns,s} + \alpha_{ns}^\T QV_{\beta \given y_s}X_s^\T] V_s^{-1} \}y_s \;  ,  \\
\text{Var}[\text{E}[\alpha^\T y \given y_s] ] &= \text{Var}[(\alpha_s^\T+ \alpha_{ns}^\T B_V  V_s^{-1})y_s] \\
&= \alpha_s^\T V_s \alpha_s + 2\alpha_{ns}^\T B_V \alpha_s + \alpha_{ns}^\T B_V V_s^{-1} B_V^\T \alpha_{ns} \;   \text{, and}\\
\text{Var}[\text{E}[\alpha^\T y \given y_s] ] &= \text{Var}[\alpha_{ns}^\T y_{ns} \given y_s] \\
&= \text{Var}[\text{E}[\alpha_{ns}^\T y_{ns} \given \beta, y_s] \given y_s] +\text{E}[\text{Var}[\alpha_{ns}^\T y_{ns} \given \beta, y_s] \given y_s]  \\
&= \text{Var}[\alpha_{ns}^\T Q\beta \given y_s]  + \text{E}[\alpha_{ns}^\T(V_{ns}- V_{ns,s}V_s^{-1}V_{s,ns})\alpha_{ns}  \given y_s] \\
&= \alpha_{ns}^\T(QV_{\beta \given y_s} Q^\T + V_{ns} - V_{ns,s}V_s^{-1}V_{s,ns})\alpha_{ns} \;.
\end{align*}

To derive the estimate given in Example \ref{ex.srs}, note that $p(\mu \given y_s) \propto N(\mu \given 0,\xi^2) \times N(y_s \given 1_n\mu, \sigma^2 I_n) \propto N(\mu \given B_{srs}b_{srs}, B_{srs})$, where $B_{srs} = (\frac{1}{\xi^2} + \frac{n}{\delta^2})^{-1}$, $b_{srs} = \frac{1}{\sigma^2} 1_n^\T y_s$, and $B_{srs}b_{srs} = \frac{\frac{1}{\sigma^2} 1_n^\T y_s} {\frac{1}{\xi^2} + \frac{n}{\sigma^2}} = \frac{\sum_{i = 1}^n y_i}{\frac{\sigma^2}{\xi^2} + n}$. Fixing the variance components, the finite population estimate is
\begin{align*}
\text{E}[\alpha^\T y \given y_s] &= \alpha_s^\T y_s + \alpha_{ns}^\T \text{E}[\text{E}[y_{ns} \given \mu, y_s] \given y_s] = \alpha_s^\T y_s + \alpha_{ns}^\T1_{(N-n)} \text{E}[\mu \given y_s] \\
&= \sum_{i=1}^n\alpha_i y_i + \frac{\sum_{i=n+1}^N \alpha_i} {\frac{\sigma^2}{\xi^2} + n} \sum_{i = 1}^n y_i 
= \sum_{i=1}^n\Bigg(\alpha_i + \frac{\sum_{i=n+1}^N \alpha_i} {\frac{\sigma^2}{\xi^2} + n} \Bigg) y_i \;.
\end{align*}

To derive (\ref{eq:ss_fp}), it is helpful to first make a note regarding $X_{s1}$ vs $X_s$ in the calculation of $p(\nu\given y_s)$ and $p(\mu \given \nu, y_s)$. Specifically, $p(\nu \given y_s)$ does not change, since $X_s = [X_{s1} : 0]$, $b = 1_n^\T X_{s1} ^\T ( X_{s1} X_{s1}^\T + \tilde{V}_s^{(\sigma)} )^{-1} y_s = 1_N^\T X_s^\T ( X_s X_s^\T + \tilde{V}_s^{(\sigma)} )^{-1} y_s$. Similarly, $B^{-1}= \frac{1}{\tilde{\gamma}^2} + 1_n ^\T X_{s1} ^\T (X_{s1} X_{s1}^\T + \tilde{V}_s^{(\sigma)} )^{-1} X_{s1} 1_n = \frac{1}{\tilde{\gamma}^2} + 1_N ^\T X_s^\T (X_s X_s^\T + \tilde{V}_s^{(\sigma)} )^{-1} X_s 1_N$.
However, while computing $p(\mu_s \given \nu, y_s)$ using $X_{s1}$ is computationally convenient for interpretation, employing $X_s$ provides us the posterior distribution $p(\mu \given \nu, y_s)$. We have that $p(\mu \given y_s, \nu) \propto N(\mu \given \nu 1_N, \delta^2 I_N) \times N(y_s \given X_s\mu, \delta^2 \tilde{V}_s^{(\sigma)} )\propto N(\mu \given B_{**}b_{**}, \delta^2 B_{**})$. Some algebra simplifies the expressions for  $b_{**}$ and $B_{**}b_{**}$ and matches the conclusions found by deriving $p(\mu_s \given \nu, y_s)$ and $p(\mu_{ns} \given \nu, y_s)$ separately:
\begin{align*}
b^{**} &= \nu 1_N + X_s^\T (\tilde{V}_s^{(\sigma)}) ^{-1} y_s
=\begin{bmatrix} \nu + \frac{\delta^2}{\sigma_1^2}  m_1 \bar{y}_1, \dots, \nu + \frac{\delta^2}{\sigma_n^2}  m_n \bar{y}_n,  \nu 1_{(N - n)}^\T \end{bmatrix} ^\T \;;\\
B_{**}^{-1} &= I_N + X_s^\T (\tilde{V}_s^{(\sigma)}) ^{-1} X_s 
=  \begin{bmatrix} \oplus_{i=1}^n \frac{\sigma_i^2 + \delta^2 m_i} {\sigma_i^2} & 0 \\ 0 & I_{(N-n)}\end{bmatrix} \;;\;
B_{**} = \begin{bmatrix}\oplus_{i=1}^N (1 - \lambda_i)\end{bmatrix} \;; \text{and} \\
B_{**}b_{**} &= \begin{bmatrix} (1 - \lambda_1) \nu + \lambda_1\bar{y}_1 ,\dots, (1 - \lambda_n) \nu + \lambda_n\bar{y}_n , \nu 1_{(N - n)}^\T \end{bmatrix} ^\T \;.
\end{align*}

Using these derivations, define $\lambda = [\lambda_1, \dots, \lambda_N]^\T$ and $\bar{y} = [\bar{y}_1, \dots, \bar{y}_n, 0_{(N-n)}]^\T$.Then fixing the variance components, we have:
\begin{align*}
\text{E}[\alpha^\T y \given y_s] &=
 \alpha_s^\T y_s + \alpha_{ns}^\T\text{E}[\text{E}[\text{E}[y_{ns} \given \mu, \nu, y_s] \given \nu, y_s]\given y_s] = \alpha_s^\T y_s + \alpha_{ns}^\T\text{E}[\text{E}[X_{ns}\mu \given \nu, y_s]\given y_s] \\
 &= \alpha_s^\T y_s + \alpha_{ns}^\T X_{ns}\text{E}[ [(1 - \lambda_1) \nu + \lambda_1\bar{y}_1,\dots, (1 - \lambda_n) \nu + \lambda_n\bar{y}_n , \nu 1_{(N - n)}^\T] ^\T\given y_s] \\
 &= \alpha_s^\T y_s + \alpha_{ns}^\T X_{ns} [\oplus_{i=1}^N (1 - \lambda_i)] 1_N \frac{\sum_{i=1}^n  \lambda_i \bar{y}_i}{1/\tilde{\gamma}^2 + \sum_{i=1}^n \lambda_i} +  \alpha_{ns}^\T X_{ns}[\oplus_{i=1}^N \lambda_i]\bar{y} \\
&= \alpha_s^\T y_s + [ \alpha_1(1 - \lambda_1), \dots, \alpha_N (1-\lambda_N)]^\T 1_N\frac{\sum_{i=1}^n  \lambda_i \bar{y}_i}{1/\tilde{\gamma}^2 + \sum_{i=1}^n \lambda_i} +  [\alpha_1 \lambda_1, \dots, \alpha_N \lambda_N]^\T\bar{y} \\
&= \sum_{i=1}^n \sum_{j=1}^{m_i} \alpha_{ij}y_{ij} + \Big\{\sum_{i=1}^N \alpha_i(1 - \lambda_i) \Big\} \frac{\sum_{i=1}^n \frac{\lambda_i}{m_i}\sum_{j=1}^{m_i} y_{ij}}{1/\tilde{\gamma}^2 + \sum_{i=1}^n \lambda_i} +  \sum_{i=1}^n \alpha_i \frac{\lambda_i}{m_i}\sum_{j=1}^{m_i} y_{ij} \\
&= \sum_{i=1}^n \sum_{j=1}^{m_i} \Bigg(\alpha_{ij} + \Bigg[\alpha_i + \frac{\sum_{i=1}^N \alpha_i (1 - \lambda_i) } {1/\tilde{\gamma}^2 + \sum_{i=1}^n \lambda_i} \Bigg]\frac{\lambda_i}{m_i} \Bigg) y_{ij} \\
  \end{align*}

Now consider the stratified case for estimating the population mean. Taking non-informative priors for the group means, $\mu$, is equivalent to letting $\delta^2 \to \infty$ and $\gamma^2 \to \infty$. Therefore $\lambda_i = \frac{\delta^2}{\delta^2 + \sigma_i^2 / m_i} \to 1$, for all $i = 1, \dots, N$. Note $\alpha_i = \sum_{j=m_i+1}^{M_i} \frac{1}{T} = \frac{M_i - m_i}{T}$, $i = 1, \dots, n$. We have that: 
\begin{align*}
\lim_{\delta^2,\gamma^2\to\infty}  \text{E}\Big[\frac{1}{T}1_T^\T y \given y_s\Big] 
&= \sum_{i=1}^n \sum_{j=1}^{m_i} \Bigg(\frac{1}{T} + \Bigg[\frac{M_i - m_i}{T}+ 0 \Bigg]\frac{1}{m_i} \Bigg) y_{ij} 
= \sum_{i=1}^n \frac{M_i}{T} \bar{y}_i 
\end{align*}

To derive (\ref{eq:sp2_fp}), note $p(\nu\given y_s, \tau^2,\Omega_s, V_s^{(\sigma)}) \propto N(y_s\given X_s1_N \nu, \delta^2 X_s X_s^\T + \Omega_s + V_s^{(\sigma)})\times N(\nu \given 0, \gamma^2) \propto N(\nu \given B_{sp2}b_{sp2}, B_{sp2})$, where $B_{sp2} = (\frac{1}{\gamma^2} + 1_N^\T X_s^\T(\delta^2 X_s X_s^\T +\Omega_s + V_s^{(\sigma)})^{-1}X_s 1_N)^{-1}$ and $b_{sp2} = 1_N^\T X_s^\T(\delta^2 X_s X_s^\T +\Omega_s + V_s^{(\sigma)})^{-1} y_s$. 
Consider the non-spatial case and define $\lambda^\T = [\lambda_1, \dots, \lambda_N] = 1_N ^\T X_s^\T (X_s X_s^\T + \tilde{V}_s^{(\sigma)} )^{-1} X_s$,  then $ B= \Big[\frac{1}{\tilde{\gamma}^2} + 1_N ^\T X_s^\T (X_s X_s^\T + \tilde{V}_s^{(\sigma)} )^{-1} X_s 1_N\Big]^{-1} =  \Big[\frac{1}{\tilde{\gamma}^2} + \lambda^\T1_N\Big]^{-1} = \Big[\frac{1}{\tilde{\gamma}^2} + \sum_{i=1}^n \lambda_i \Big]^{-1}$, which agrees with our previous findings. 

Similarly, define $\lambda^{*\T} = [\lambda_1^*, \dots, \lambda_N^*] = 1_N^\T X_s^\T(\delta^2 X_s X_s^\T +\Omega_s + V_s^{(\sigma)})^{-1}X_s$. 

Then $B_{sp2} = \Big[\frac{1}{\tilde{\gamma}^2} + \lambda^{*\T}1_N\Big]^{-1} = \Big[\frac{1}{\tilde{\gamma}^2} + \sum_{i=1}^n \lambda_i^*\Big]^{-1}$.

Additionally, $p(\mu \given y_s, \nu) \propto N(\mu \given \nu 1_N, \delta^2 I_N) \times N(y_s \given X_s\mu, \Omega_s + V_s^{(\sigma)})\propto N(\mu \given B_{sp2*}b_{sp2*}, B_{sp2*})$. Here $B_{sp2*} = (\frac{1}{\delta^2} I_N + X_s^\T (\Omega_s + V_s^{(\sigma)})^{-1}X_s)^{-1}$ and $b_{sp2*} = \frac{1}{\delta^2}1_N\nu + X_s^\T(\Omega_s + V_s^{(\sigma)})^{-1}y_s$.

Fixing the variance parameters, we have that:
\begin{align*}
&\text{E}[\alpha^\T y \given y_s] = \alpha_s^\T y_s +  \alpha_{ns}^\T\text{E}[ \text{E}[\text{E}[y_{ns} \given \mu, \nu, y_s]\given \nu, y_s] \given y_s]\\
&= \alpha_s^\T y_s +  \alpha_{ns}^\T\text{E}[\text{E}[X_{ns} \mu + \Omega_{ns,s}(\Omega_s + V_s^{(\sigma)})^{-1}(y_s - X_s \mu)\given \nu, y_s]\given y_s]\\
&= \alpha_s^\T y_s +  \alpha_{ns}^\T\Omega_{ns,s}(\Omega_s + \sigma^2V_s^{(\sigma)})^{-1}y_s + \alpha_{ns}^\T[X_{ns} - \Omega_{ns,s}(\Omega_s + V_s^{(\sigma)})^{-1}X_s] \times \\
& \quad \quad \text{E}\begin{bmatrix}(\frac{1}{\delta^2} I_N + X_s^\T (\Omega_s + V_s^{(\sigma)})^{-1}X_s)^{-1} (\frac{1}{\delta^2}1_N\nu + X_s^\T(\Omega_s + V_s^{(\sigma)})^{-1}y_s) \given y_s\end{bmatrix}\\
&= \alpha_s^\T y_s +  \alpha_{ns}^\T\Omega_{ns,s}(\Omega_s + \sigma^2V_s^{(\sigma)})^{-1}y_s + \alpha_{ns}^\T[X_{ns} - \Omega_{ns,s}(\Omega_s + V_s^{(\sigma)})^{-1}X_s] \times \\
& \quad \quad \begin{pmatrix}\frac{1}{\delta^2} I_N + X_s^\T (\Omega_s + V_s^{(\sigma)})^{-1}X_s\end{pmatrix}^{-1} X_s^\T(\Omega_s + V_s^{(\sigma)})^{-1}y_s + \\
&\quad \quad \alpha_{ns}^\T[X_{ns} - \Omega_{ns,s}(\Omega_s + V_s^{(\sigma)})^{-1}X_s] \times \\
& \quad \quad \begin{pmatrix}\frac{1}{\delta^2} I_N + X_s^\T (\Omega_s + V_s^{(\sigma)})^{-1}X_s\end{pmatrix}^{-1} \frac{1}{\delta^2} 1_N \frac {1_N^\T X_s^\T(\delta^2 X_s X_s^\T +\Omega_s + V_s^{(\sigma)})^{-1} y_s}{\frac{1}{\gamma^2} + \sum_{i=1}^n \lambda_i^*} \\
&= \sum_{i=1}^n \sum_{j=1}^{m_i} \Bigg(\alpha_{ij}+  \alpha_{ns}^\T\Bigg\{\Omega_{ns,s}(\Omega_s + V_s^{(\sigma)})+ \\
&[X_{ns} - \Omega_{ns,s}(\Omega_s + V_s^{(\sigma)})^{-1}X_s]
 \begin{pmatrix}\frac{1}{\delta^2} I_N + X_s^\T (\Omega_s + V_s^{(\sigma)})^{-1}X_s\end{pmatrix}^{-1} \times \\&\Bigg[X_s^\T (\Omega_s + V_s^{(\sigma)})^{-1} +  \frac {\frac{1}{\delta^2} 1_N1_N^\T X_s^\T(\delta^2 X_s X_s^\T +\Omega_s + V_s^{(\sigma)})^{-1}}{\frac{1}{\gamma^2} + \sum_{i=1}^n \lambda_i^*}\Bigg]\Bigg\}q_{ij}\Bigg)y_{ij} \;.
\end{align*}

\section*{Appendix B}

To perform the two-stage procedure implemented in Section~\ref{sim.exact}, sample means, $\hat{\mu}_i$, and sample variances, $\hat{\sigma}_i^2$, were calculated from each observed region, $i = 1, \dots, 25$. The variance matrix of the sampled units was fixed to be $\tilde{V}_s^{(\sigma)}  = \begin{bmatrix} \oplus_{i = 1}^n \frac{\hat{\sigma}_i^2}{\text{Var}(\hat{\mu})} I_{m_i}\end{bmatrix}$where $\text{Var}(\hat{\mu})$ represents the sample variance of the observed sample means. Similar to fixing $V_s^{(\sigma)}$, the variance matrix of the nonsampled units was fixed at $V_{ns}= \begin{bmatrix} \oplus_{i = 1}^N \frac{\tilde{\sigma}_i^2}{\text{Var}(\hat{\mu})} I_{M_i - m_i}\end{bmatrix}$, where $\tilde{\sigma}_i^2 = \hat{\sigma}_i^2$ if $i \in \{1, \dots, n\}$ and $\tilde{\sigma}_i^2 = \frac{1}{n}\sum_{i=1}^n \hat{\sigma}_i^2$ if $i \in \{n+1, \dots, N\}$. The value of $\gamma^2$ was fixed to be half of the value of $\delta^2$, reflecting the belief that there was less variability in the population mean than between group means. The prior distribution for $\delta^2$ was assigned to be $IG(3,5)$. Sampling from the posterior was performed using the conditional distributions and methods described in Section \ref{sec:exact_models}. As we have fixed the ratios of all variance components, we have also fixed $\tilde{\lambda}_i = \text{Var}(\hat{\mu})/(\text{Var}(\hat{\mu})+ \hat{\sigma}_i^2/m_i)$ if $i \in (1, \dots, n)$ and $\tilde{\lambda}_i = 0$ if $i \in (n+1, \dots, N). $.  Define $c = \frac{\sum_{i=1}^n  \tilde{\lambda}_i \bar{y}_i}{\frac{1}{2} + \sum_{i=1}^n \tilde{\lambda}_i }$, $d = \Big[\frac{1}{2} + \sum_{i=1}^n \tilde{\lambda}_i\Big]^{-1}$, $c^{*(g)} = \begin{bmatrix} (1 - \tilde{\lambda}_1) \nu^{(g)} + \tilde{\lambda}_1\bar{y}_1 \\ \vdots \\ (1 - \tilde{\lambda}_n) \nu^{(g)} + \tilde{\lambda}_n\bar{y}_n \end{bmatrix}$, and $d^* = \begin{bmatrix} \oplus_{i=1}^n (1 - \tilde{\lambda}_i) \end{bmatrix}$. The following procedure was implemented to produce posterior estimates of the population mean, $\bar{y}^{(g)}$, for $G$ iterations. 

\begin{align*}
\text{for(g in 1:G)}\{ \quad \quad & \\
\delta^{2 (g)} &\sim IG\Bigg(3+\frac{1}{2}\sum_{i=1}^n m_i, 5 + \frac{1}{2}\Big[y_s^\T(\tilde{V}_s^{(\sigma)}  + X_sX_s^\T)^{-1}y_s + \frac{c^2}{d}\Big] \Bigg)\\
\nu^{(g)} &\sim N(c,\delta^{2(g)}d) \\
\mu_s^{(g)} &\sim N(c^{*(g)}, \delta^{2(g)}d^*) \\
\mu_{ns}^{(g)} &\sim N(\nu^{(g)}1_n, \delta^{2(g)} I_{N-n}) \\
y_{ns}^{(g)} &\sim N(X_{ns}\mu^{(g)}, \delta^{2(g)}\tilde{V}_{ns}) \\
\bar{y}^{(g)} &= \frac{1}{T}\Big(\sum_{i = 1}^n\sum_{j=1}^{m_i} y_{ij} + \sum_{i = 1}^N\sum_{j=m_i + 1}^{M_i} y_{ij}^{(g)}\Big) \\
 \} \quad &
\end{align*}

Figure~\ref{fig:sim3} recreates the centered mean plots presented in Figure~\ref{fig:sim2} for the larger data case, in which the number of regions is 324. As in the $N=100$ case, the point estimates and 95\% credible intervals are similar for the two models. Additionally, the regional spatial model still outperforms the two-stage model with a WAIC of 1,052 (SE = 38.89) compared to 1,869 (SE = 34.77). MCMC sampling was performed using JAGS \citep{Plummer:2017vp} in R \citep{Rlang}.

\begin{figure}[ht]
\centering
\includegraphics[width = 4in]{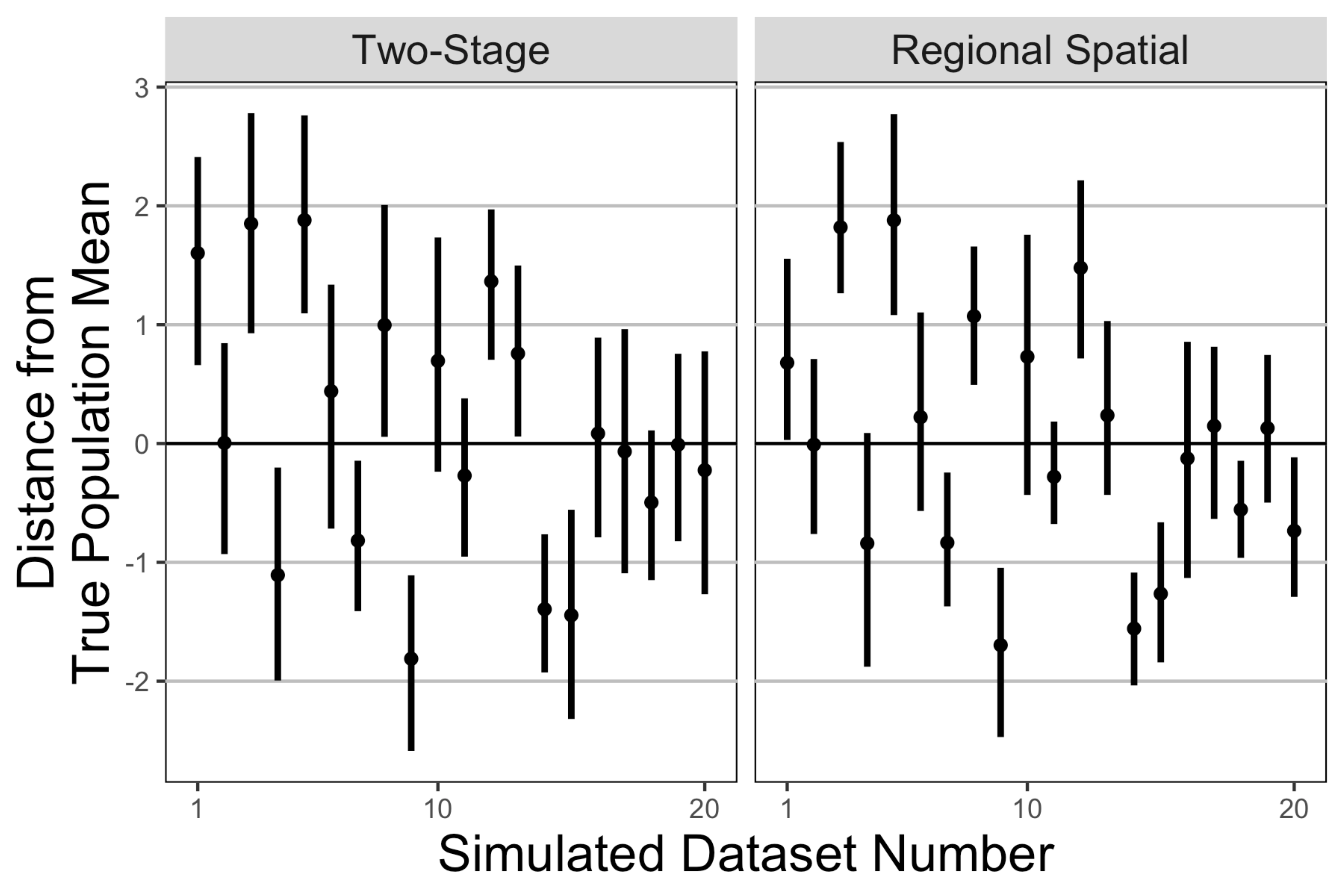}
\caption{Centered Population Mean Estimates from 2 MCMC Models with 95\% CI.}
\label{fig:sim3}
\end{figure}

Table~\ref{tab:sim2} compares the finite population mean estimate, $\nu$ estimate, and model fit for the first simulated dataset. Recall that $\nu = 2$ and the finite population mean was 2.60. Notice that while the true values are included in the credible intervals for all models, the credible intervals for $\nu$ are wider than those of the FP mean.
\begin{table}[!ht]
\caption{Comparison of Estimates and Model Fit from MCMC }
\centering
\begin{tabular}{llll}
  \hline
 Model & FP Mean (95\% CI) & $\nu$ (95\% CI) & WAIC (SE) \\ 
  \hline
  1. Two-Stage & 2.84 (1.93, 3.67) & 2.84 (1.82, 3.82) & 1870.26 (35.00) \\ 
  2. Spatial & 3.04 (1.33, 4.56)  & 3.24 (1.20, 5.01) & 1912.70 (26.36)  \\ 
  3. Two-Stage + Spatial & 3.11 (-0.86, 6.70) & 3.22 (-1.23, 7.33) & 455.67 (17.03)  \\ 
  4. Regional Spatial & 2.56 (1.74, 3.37)& 2.44 (1.49, 3.47) & 1202.08 (38.99) \\ 
   \hline
\end{tabular}
\bigskip
\label{tab:sim2}
\end{table}

\section*{Appendix C}

Figure~\ref{fig:zip} presents the 2010 zip code tabulation areas in the California Central Valley obtained from the \textit{tigris} R package \citep{Walker:2018ti}. Sixty-three zip codes were included in the analysis of a subset of observations from the UC Davis Nitrate Data in the data repository of the \cite{GAMA}, presented in Section~\ref{sec:data_analysis}. These zip codes are denoted as either sampled or non-sampled, while all other zip codes are denoted as excluded.
\begin{figure}[!ht]
\centering
\includegraphics[width = 5in]{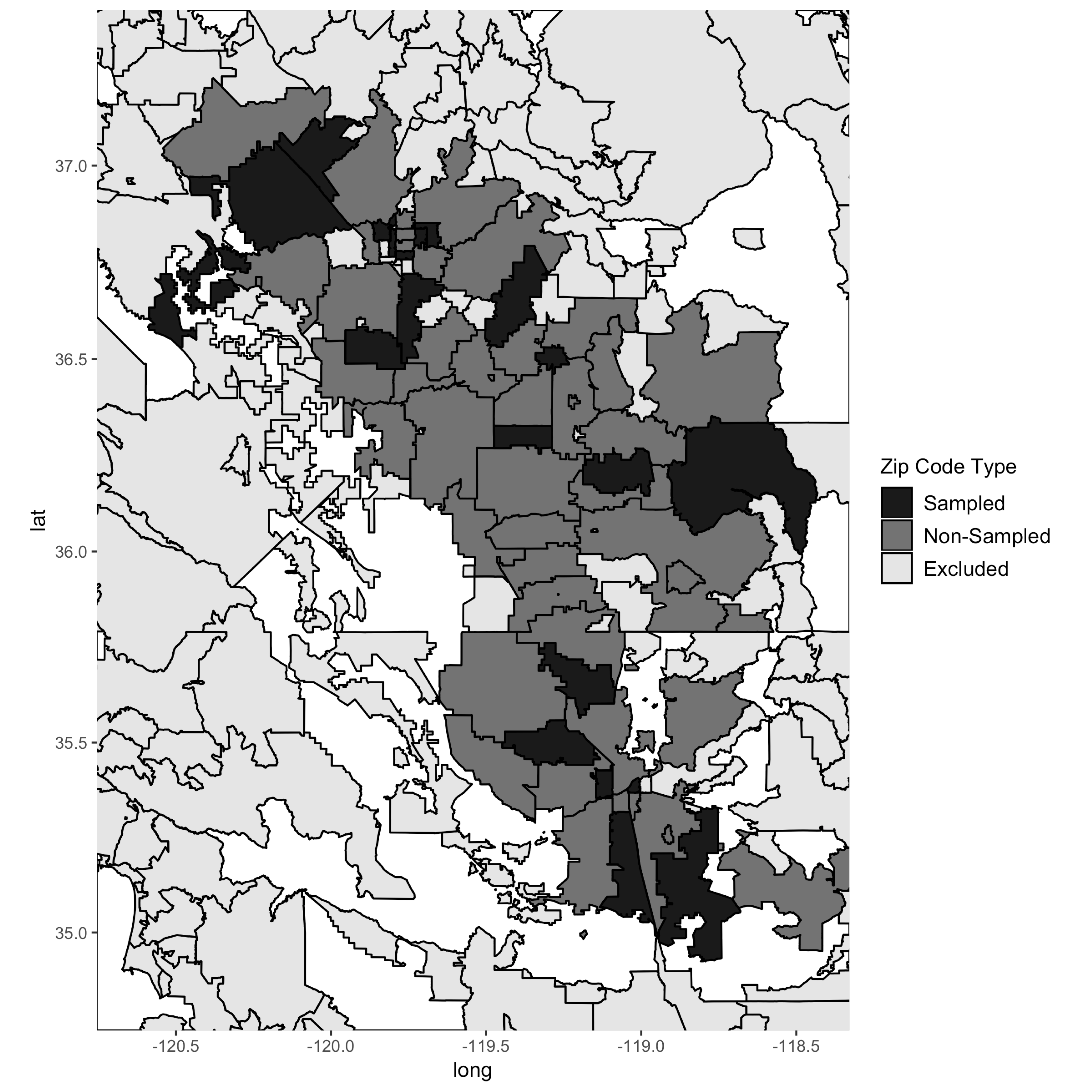}
\caption{Plot of California zip code tabulation areas.}
\label{fig:zip}
\end{figure}

The variograms provided in Figure~\ref{fig:davario} were fitted to the entire population of wells (left) and sampled wells (right). For the population variogram, the estimated values of the nugget, partial sill, and range were 2527.4, 319.9, and 2.29, respectively. For the sample variogram, the estimated values of the nugget, partial sill, and range were 1808.6, 1014.4, and 335.9, respectively. 

\begin{figure}[!ht]
\centering
\begin{minipage}{.5\textwidth}
  \centering
  \includegraphics[width=.5\linewidth]{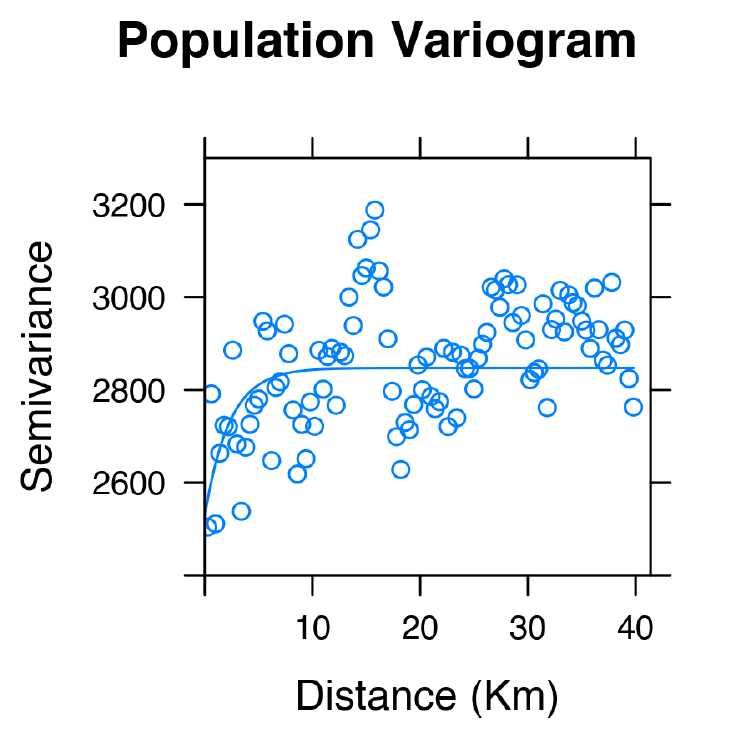}
\end{minipage}%
\begin{minipage}{.5\textwidth}
  \centering
  \includegraphics[width=.5\linewidth]{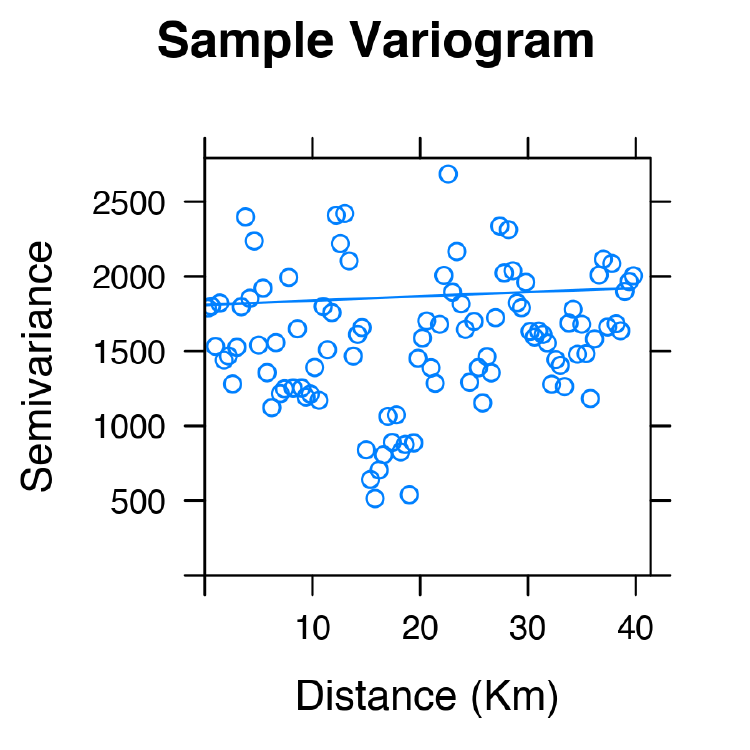}
\end{minipage}
  \caption{Variograms from Population and Sampled Data.}
    \label{fig:davario}
\end{figure}

Figure~\ref{fig:da2} provides spatial residual plots arising from the three spatial models. The spatial model which does not account for regional effects sees the most dispersed spatial effects, while the two-stage + spatial and regional spatial models show more localized spatial variability. MCMC sampling was performed using JAGS \citep{Plummer:2017vp} in R \citep{Rlang}.

\begin{figure}[!ht]
\centering
\includegraphics[width = \textwidth]{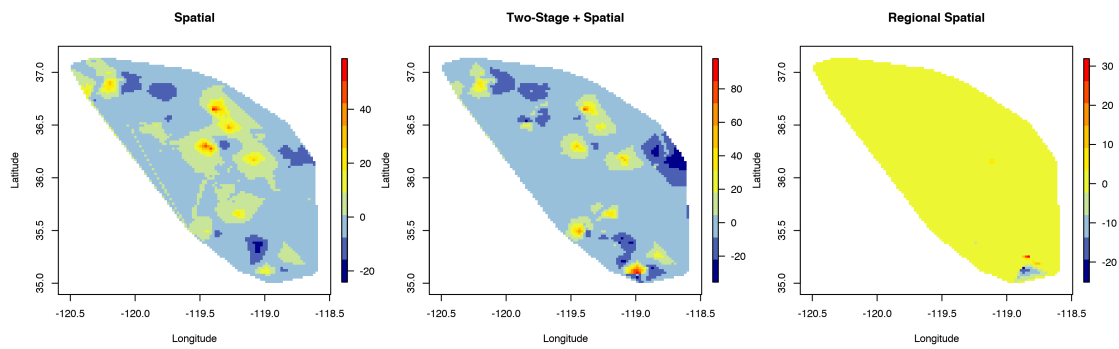}
\caption{Spatial residual plots from the three spatial models.}
\label{fig:da2}
\end{figure}

\end{document}